\documentclass[12pt]{article}
\pdfoutput=1
\usepackage{jheppub}

\usepackage{amsmath,amssymb,amsfonts}
\usepackage{graphicx}
\usepackage{color}
\usepackage{tikz}
\usepackage{dsfont}

\usepackage{booktabs}
\usepackage{multirow}

\usepackage{subfigure}

\usepackage{comment}

\makeatletter\def\l@subsubsection#1#2{}%
\makeatother

\def\be{\begin{eqnarray}}
\def\ee{\end{eqnarray}}

\def\makeatletter{\catcode`\@=11}
\makeatletter
\def\mathbox#1{\hbox{$\m@th#1$}}%
\def\math@ccstyles#1#2#3#4#5#6#7{{\leavevmode
      \setbox0\mathbox{#6#7}%
      \setbox2\mathbox{#4#5}%
      \dimen@ #3%
      \baselineskip\z@\lineskiplimit#1\lineskip\z@
      \vbox{\ialign{##\crcr
             \hfil \kern #2\box2 \hfil\crcr
             \noalign{\kern\dimen@}%
             \hfil\box0\hfil\crcr}}}}
\def\mathaccstyles{\math@ccstyles\maxdimen}
\def\maththroughstyles{\math@ccstyles{-\maxdimen}}
\def\unity%
 {\maththroughstyles{.45\ht0}\z@\displaystyle {\mathchar"006C}\displaystyle 1}

\begin{document}

\title{The Cat's Cradle\footnote{The \href{https://en.wikipedia.org/wiki/Cat\%27s\_cradle}{cat's cradle} is a sequence game in which two or more players use
a loop of string to form web patterns that change at each step.}: Deforming the higher rank $E_1$ and $\tilde{E}_1$ theories}

\author[a]{Oren Bergman,}
\emailAdd{bergman@physics.technion.ac.il}
\author[b]{Diego Rodr\'iguez-G\'omez,}
\emailAdd{d.rodriguez.gomez@uniovi.es}

\affiliation[a]{Department of Physics, Technion, Israel Institute of Technology\\
Haifa, 32000, Israel\\[-4mm]}

\affiliation[b]{Department of Physics, Universidad de Oviedo\\
Avda.~Calvo Sotelo 18, 33007, Oviedo, Spain\\[-4mm]
}

\abstract{We use 5-brane webs to study the two-dimensional space of supersymmetric mass deformations of
higher rank generalizations of the 5d $E_1$ and $\tilde{E}_1$ theories.
Some of the resulting IR phases are described by IR free supersymmetric gauge theories,
while others correspond to interacting fixed points.
The number of different phases appears to grow with the rank.
The space of deformations is qualitatively different for the even and odd rank cases, 
but that of the even (odd) rank $E_1$ theory is similar to that of the odd (even) rank
$\tilde{E}_1$ theory. 
One result of our analysis predicts that the supersymmetric $SU(N)$ theory with CS level 
$k=\frac{N}{2} + 4$ and 
a single massless antisymmetric hypermultiplet exhibits an enhanced global symmetry at the UV
fixed point,
given by $SU(2)\times SU(2)$ if $N$ is even, and $SU(2)\times U(1)$ if
$N$ is odd.}

\maketitle

\section{Introduction}

The existence of UV complete supersymmetric quantum field theories in five 
spacetime dimensions,
first demonstrated by Seiberg in \cite{Seiberg:1996bd},  
can be regarded as one of the most striking predictions of string theory.\footnote{The same can be said 
about supersymmetric quantum field theories in six spacetime dimensions.}
Superconformal fixed points in five dimensions were originally argued to exist by observing the behavior of 
minimally-supersymmetric 5d gauge theories on the Coulomb branch \cite{Seiberg:1996bd,Morrison:1996xf,Intriligator:1997pq}.
Under certain conditions the effective coupling squared remains positive and finite everywhere on the Coulomb branch,
and this indicates the possibility of an interacting UV fixed point at the origin.
The 5d gauge theory is the result of deforming the 5d superconformal field theory (SCFT)
by a relevant mass parameter and flowing to the IR.
The mass parameter in this case becomes the inverse-squared-YM-coupling of the gauge theory.
Further recent explorations using the tools of string theory have made it apparent that the 
space of 5d SCFT's is in fact much larger,
and includes gauge theories that do not satisfy the aforementioned requirement,
as well as a plethora of theories that cannot be deformed into a gauge theory.
There are a number of different approaches by now to the construction of 5d SCFT's,
including 5-brane webs in Type IIB string theory \cite{Aharony:1997ju,Aharony:1997bh,Bergman:2015dpa,Zafrir:2015ftn,Hayashi:2018bkd,Hayashi:2018lyv,Hayashi:2019yxj}, 
geometric engineering in M-theory \cite{Douglas:1996xp,Xie:2017pfl,Jefferson:2018irk,Closset:2018bjz,Apruzzi:2019vpe,
Apruzzi:2019opn,Apruzzi:2019enx,Apruzzi:2019kgb,Closset:2020scj},
and reduction of 6d theories 
\cite{Bhardwaj:2018yhy,Bhardwaj:2018vuu,Bhardwaj:2019jtr,Bhardwaj:2019xeg,Bhardwaj:2020gyu,Bhardwaj:2020ruf,Bhardwaj:2020avz}.
It seems however fair to say that the precise connection between all these approaches and whether this fully exhausts the space of 5d SCFT's is yet to be clarified.

An interesting set of questions one can address within these constructions
is related to the spaces of supersymmetric deformations of five-dimensional SCFT's.
In particular 5d SCFT's generally have a moduli space containing both Coulomb and 
Higgs branches. 
In theories with an IR gauge theory description these correspond
to vacuum expectation values of scalars in vector multiplets and hypermutiplets, respectively.
The Higgs branch is especially interesting since it is generally richer than 
what appears in the IR gauge theory.
There have been a number of recent investigations of Higgs branches motivated by and using
string theory constructions \cite{Ferlito:2017xdq,Cabrera:2018jxt,Bourget:2019rtl,Bourget:2020gzi,Akhond:2020vhc,vanBeest:2020kou}.

One can also consider deforming 5d SCFT's by relevant operators. 
Five-dimensional SCFT's have no marignal operators preserving supersymmetry,
and the only relevant operators correspond to mass parameters.
These are dimension four scalar operators that sit inside the conserved current supermultiplets
associated to the global symmetry of the theory.
The mass itself may therefore be regarded as the VEV of a scalar field in a background
vector multiplet associated to the global symmetry.
The number of independent supersymmetric mass deformations is therefore equal to the rank
of the global symmetry.
In some cases a mass deformation of a 5d SCFT leads to an IR free supersymmetric gauge theory,
where the value of the mass becomes the inverse-squared-YM-coupling of the gauge theory.
But more generally a mass deformation may also lead to another interacting fixed point in the IR.
Furthermore, since the mass parameter in five dimensions is real, the theory may flow 
to different IR phases for positive and negative mass, and each of these may be an IR 
free gauge theory or an interacting theory.

The simplest non-trivial examples of this are the $E_1$ and $\tilde{E}_1$ theories \cite{Seiberg:1996bd,Morrison:1996xf}.
Both have a rank one global symmetry, $SU(2)$ in the first case and $U(1)$ in the second, and therefore
a single supersymmetric mass parameter.
The deformation of the $E_1$ theory leads to a supersymmetric $SU(2)$ gauge theory 
with a trivial theta parameter on both sides,
whereas the deformation of the $\tilde{E}_1$ theory leads to a supersymmetric $SU(2)$ gauge theory 
with a non-trivial theta parameter on one side, and to the interacting $E_0$ theory on the other side.

When there are several mass parameters the situation gets more interesting, and the space of mass deformations
can potentially have many different phases separated by critical walls.
A simple example of this is the $E_2$ theory, which has a two-dimensional space of 
supersymmetric mass deformations exhibiting three different phases \cite{Morrison:1996xf}.
We will recall this example and its string theory description in section 2.

In this paper we will explore the two-dimensional space of supersymmetric mass
deformations of another set of theories describing higher rank generalizations of the $E_1$ and 
$\tilde{E}_1$ theories.
Unlike the $E_2$ case in \cite{Morrison:1996xf}, we will not be able to fully map out 
the space of deformations using field theory alone.
We will therefore mainly use the realization of these theories using 5-brane webs in Type IIB
string theory.
The mass deformations of the field theories will be realized as geometric deformations 
of the 5-brane webs.
These deformations give rise to various junction splitting and joining transitions,
resembling the changing patterns of the cat's cradle. 
As we will see, the number of different phases and critical lines appears to grow with increasing rank,
and the nature of the phases is qualitatively different in the even and odd rank cases.
Some of the phases are IR free gauge theories while others are interacting theories.

The rest of this paper is organized as follows.
In section 2 we will briefly recall the different phases of the $E_2$ theory,
and how they are are realized in the 5-brane web description.
In section 3 we will study the deformations of higher rank $E_1$ theories,
and in section 4 we will study those of the 
higher rank $\tilde{E}_1$ theories.
Section 5 contains our conclusions.
We also include an important Appendix, in which we review the basics of 5-brane webs
and their deformations.

\section{Warm-up: the $E_2$ theory}

\begin{figure}[h!]
\center
\includegraphics[height=0.3\textwidth]{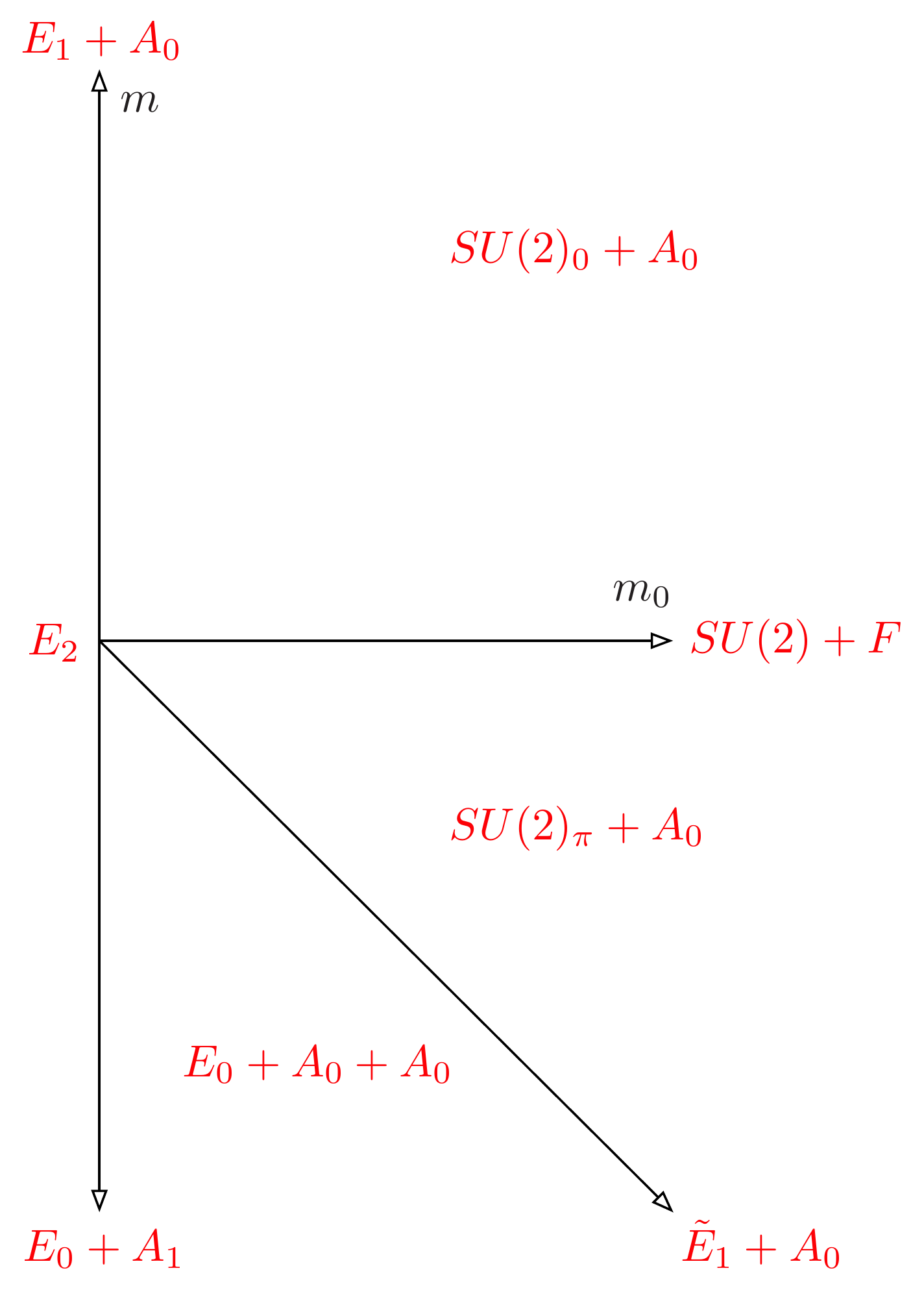} 
\caption{Phases of the $E_2$ theory \cite{Morrison:1996xf}.}
\label{E2Phases}
\end{figure}

The $E_2$ theory is part of the series of rank 1 interacting 5d SCFTs that have an $E_n$ global symmetry,
with $n=1,\ldots , 8$, introduced in \cite{Seiberg:1996bd}.
This particular theory has a global symmetry $E_2=SU(2)\times U(1)$, and 
correspondingly, a two-dimensional space of supersymmetric mass deformations, 
one for each Cartan factor of the global symmetry group.
This space was mapped out in \cite{Morrison:1996xf}, and is reproduced in Fig.~\ref{E2Phases}.
The mass $m_0$ (the $x$-axis) corresponds to the $SU(2)$ factor and the mass $m$ (the $y$-axis) corresponds 
to the $U(1)$ factor. Since $m_0$ can be mapped to $-m_0$ by an $SU(2)$ transformation, it suffices
to present just the right hand side of the deformation plane, $m_0\geq 0$. 
The left hand side is its mirror image.

Let us begin with the $x$-axis, namely the line $m=0$.
Along this line the theory flows in the IR to an $SU(2)$ gauge theory with a 
single flavor hypermultiplet,
denoted $SU(2)+F$, and the mass $m_0>0$ corresponds to the inverse-squared-YM-coupling.
The global symmetry of this theory is $U(1) \times U(1)$, where one $U(1)$ factor is the
flavor symmetry, and the other $U(1)$ factor is the topological symmetry.
This is consistent with the fact that $m_0$ is part of a triplet of the global $SU(2)$ symmetry,
and therefore breaks $SU(2) \rightarrow U(1)$.
Going above or below this line corresponds to turning on a positive or negative mass
$m$ for the flavor. 
The massive flavor decouples from the low energy theory, and its only effect is a mod 2
theta parameter whose value depends on the sign of the mass.
Above the $m_0$ axis the low energy theory is a pure $SU(2)_0$ gauge theory, 
and below the $m_0$ axis it is a pure $SU(2)_\pi$ gauge theory.
While all of this is true at the origin of the Coulomb branch,
there is a point on the Coulomb branch, roughly when $\phi = |m|$,
where the flavor becomes massless, giving rise to an extra sector in the low energy theory. It is useful to keep track of this sector.
Morrison and Seiberg denoted this sector by $A_0$, where $A_n$ denotes
a free $U(1)$ gauge theory with $n+1$ equally charged hypermultiplets \cite{Morrison:1996xf}.

In the upper-half plane no new phase is encountered until one reaches the positive $y$ axis, 
namely the line $m_0=0$.
This corresponds to the infinite effective coupling limit of the $SU(2)_0$ theory, and is described by
the $E_1$ theory. This has a global $E_1 = SU(2)$ symmetry, consistent with its position along $m_0=0$.
The massive flavor sector $A_0$ goes along for the ride. Taking its mass to zero brings us back to
the $E_2$ theory.

The lower half-plane is more interesting.
The effective coupling of the $SU(2)_\pi$ theory blows up along the line $m_0 = -4m$.
Along this line the theory flows to the $\tilde{E}_1$ theory, which has only a $U(1)$ global symmetry.
The massive sector $A_0$ again goes along for the ride.
On the other side of this line the theory flows to a different IR fixed point, described
by the $E_0$ theory, a rank one interacting theory with no global symmetry,
together with two $A_0$ sectors.
The second massive field is, roughly speaking, the $SU(2)$ instanton.
Along the negative $y$ axis we expect to recover an $SU(2)$ global symmetry.
This happens by the merger of the two $A_0$ sectors into an $A_1$ \cite{Morrison:1996xf}.

\begin{figure}[h!]
\center
\includegraphics[height=0.5\textwidth]{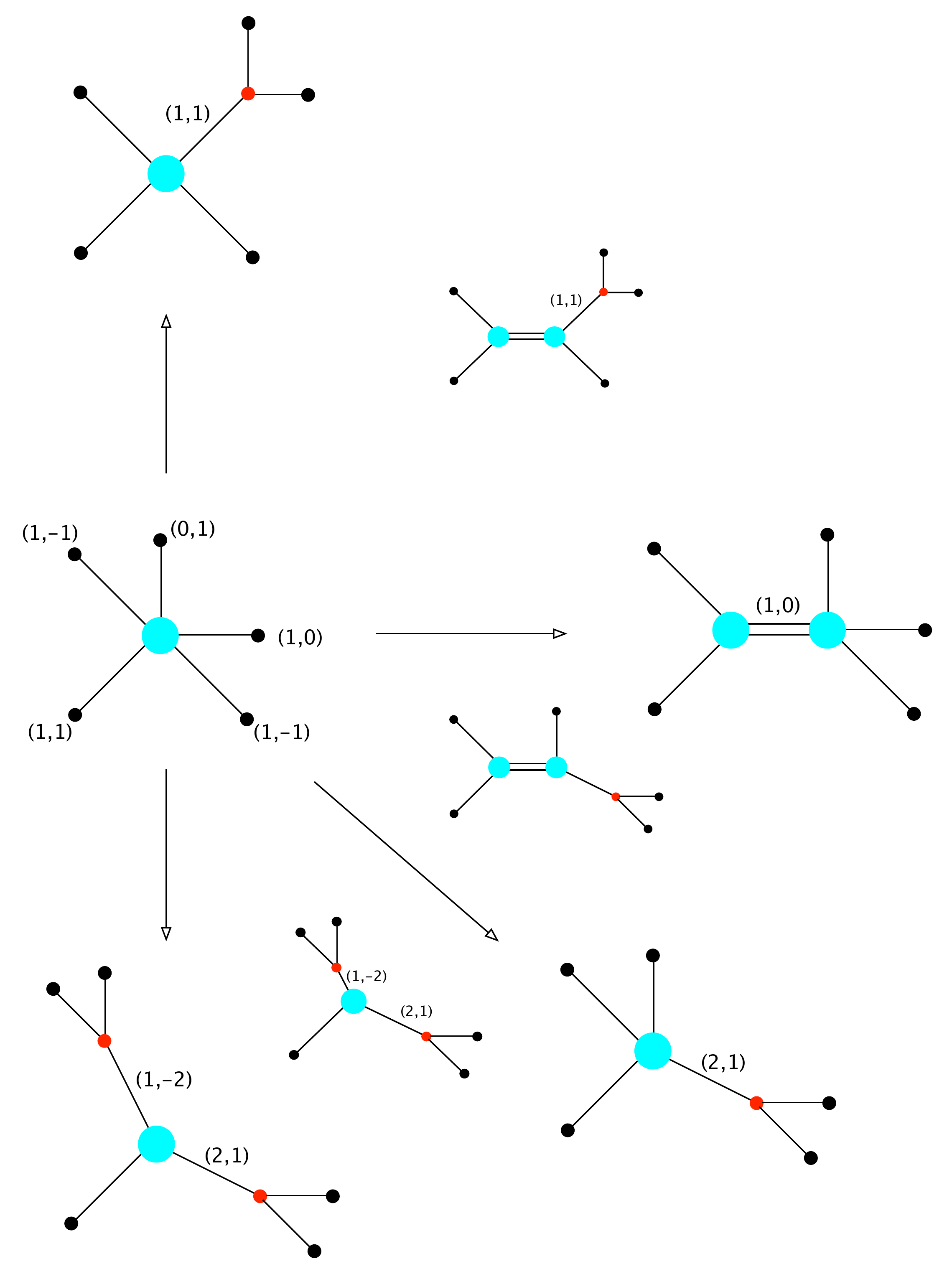} 
\caption{The 5-brane webs corresponding to the phases of tyhe $E_2$ theory in Fig.~\ref{E2Phases}.}
\label{E2Webs}
\end{figure}

This entire structure is beautifully reproduced using 5-brane webs in Type IIB string theory \cite{Aharony:1997ju}. In Fig.~\ref{E2Webs} we have reproduced Fig.~11 of \cite{Aharony:1997ju},
with the added ingredients describing the massive sectors.
The latter correspond to the {\em trivial junctions}, denoted by red circles, 
that are equivalent to detached 7-branes (see the Appendix for more details).

\section{The ${E}_1$ theory}

The $E_1$ theory is one of the simplest interacting superconformal field theories in five dimensions.
It is a rank one theory with an $E_1=SU(2)$ global symmetry, and hence admits a single real mass deformation. 
Deforming the theory with this mass in either direction leads to the IR free supersymmetric gauge theory with gauge group $Sp(1)_0=SU(2)_0$. 
As discussed above, the fact that the positive and negative mass deformations lead to the same IR theory is guaranteed by the global $SU(2)$ symmetry 
of the massless theory. 
The mass deformation breaks $SU(2)$ to $U(1)$, which is in turn realized as the topological symmetry of the five-dimensional $SU(2)$ gauge theory.

The existence of the $E_1$ theory emerged originally from a string theory construction involving a D4-brane probing an orientifold 8-plane in Type IIA string theory. This set-up allows a natural generalization for $N$ D4-branes. We will denote the low energy theory on the D4-branes as ${E}_1^{(N)}$. By construction this is the UV fixed point of a supersymmetric gauge theory with gauge group $Sp(N)_0$ and one hypermultiplet in the two-index antisymmetric representation. For $N=1$ this is a singlet, and the gauge theory is effectively a pure supersymmetric $Sp(1)_0=SU(2)_0$ theory.
For $N>1$ this theory has an $SU(2)\times SU(2)$ global symmetry, and correspondingly a two-dimensional space of supersymmetric real mass deformations.
We will denote the coordinates of this space by $(m_0,m)$.
The $SU(2)\times SU(2)$ global symmetry implies that the four quadrants are identical, and allows us to focus on just one quadrant, which we will take to be $m_0,m\geq 0$. 
Turning on just $m_0$, the theory flows to the $Sp(N)_0 + AS$ gauge theory, where $m_0$ is interpreted as the inverse-squared-YM coupling. The first $SU(2)$ factor breaks to $U(1)$, which is realized as the topological symmetry of the gauge theory, and the second $SU(2)$ remains as the matter symmetry acting on the real antisymmetric hypermultiplet. Then, deforming along $m$ gives a mass to the antisymmetric hypermultiplet, resulting in a pure $Sp(N)_0$ gauge theory in the IR.
From this point of view, the reason that the positive and negative $m$ deformations are identical
is that the effect of the antisymmetric fermion on the theta parameter is doubled relative 
to a fundamental fermion.

The VEV of the scalar in the vector multiplet also contributes to the mass
of the hypermultiplet, and, as in the case of the $E_2$ theory, a part of it will become
massless at some point on the Coulomb branch.
In particular in the direction where $Sp(N)$ is broken to $Sp(N-1)\times U(1)$ there is 
a $U(1)$ charged state in the fundamental representation of $Sp(N-1)$ which is massless
for $\phi = m$. This is again an $A_0$ sector, and the 
full IR theory is denoted $Sp(N)_0 + A_0$.\footnote{Using \cite{Bergshoeff:2002qk,Bergshoeff:2004kh}, it is easy to see that the scalar potential is 
\begin{equation}
\label{V}
V=\frac{P^I\,P^I}{m_0}+\frac{1}{2}\,(\phi-m)^2\, (a^i)^{\dagger}\,a^i+2\,m\,\phi\, (a^i)^{\dagger}\,\mathcal{P}^i_j\,a^j\,,\qquad \mathcal{P}^i_j=\frac{\unity\,\delta^i_j-t\, (J_{3})^j_i}{2}\,,
\end{equation}
where we combine the 4 scalars in the antisymmetric hypermultiplet into two complex combinations, denoted by $a_i$, $i=1,\,2$, which form an $SU(2)_R$ doublet. Moreover, we denote by $P^I\equiv P^I_{\alpha}$ the moment maps of the $Sp(N)$ gauge action being $I$ an $SU(2)_R$ triplet index --here $J_I$ are the $SU(2)_R$ generators-- and $\alpha$ an $Sp(N)$ index --with $t$ the $Sp(N)$ generator along the chosen Coulomb branch direction respectively. Finally, we note that $\mathcal{P}$ --for which we only explicitly display the $SU(2)_R$ indices, is a projector, \textit{i.e.} $\mathcal{P}^2=\unity$. For $\phi\ne m$ the whole $AS$ becomes massive leaving the pure $Sp(N)_0$ theory. On top of it, at $\phi=m$, the part of the $AS$ satisfying $\mathcal{P}a=0$ remains massless: this is the $A_0$ theory.}

By analogy with what we saw in the previous section, this might lead one to guess that turning on just $m$ would lead in the IR to the SCFT corresponding to the UV fixed point of the pure $Sp(N)_0$ gauge theory.
However this cannot be correct, since this theory is known to have only a $U(1)$ global symmetry \cite{Zafrir:2015uaa}, whereas the theory with $m_0=0$ should have an unbroken $SU(2)$ global symmetry.
This implies that there are necessarily more phases.

To gain further insight, we may consider the gauge theory, which, for $m_0\gg m$, is weakly coupled. We can use the perturbative result for the effective prepotential on the Coulomb branch, which reads (in the chamber $m_0,\,m>\phi_N\geq \phi_{N-1}\geq \cdots \geq \phi_1$)
\begin{equation}
\label{FIMS}
\mathcal{F}(\phi_i)=\frac{1}{3}\sum_{i=1}^N\phi_i^3+\sum_{i=1}^N\sum_{j=1}^N\phi_i^2\phi_j+\big(m_0-m\,(N-1)\big)\,\sum_{i=1}^N\phi_i^2-\frac{N\,(N-1)}{6}\,m^3\,;
\end{equation}
and from this, compute the effective YM coupling at the origin
\begin{equation}
\label{couplingattheorigin}
\frac{1}{g_{eff}^2} = 2\,\Big(m_0-(N-1) m\Big) \,.
\end{equation}
This exhibits a critical line given by $m_0=(N-1) m$. 
In the rank one case this reduces to $m_0=0$, corresponding to the infinite coupling limit
of the pure $SU(2)_0$ theory. But for higher rank it is a diagonal line with a slope $1/(N-1)$.

The perturbative analysis of the $Sp(N)_0 + AS$ gauge theory breaks down at this critical line.
To proceed further we will use the Type IIB string theory embedding of the $E_1^{(N)}$ theory
in terms of 5-brane webs. This will allow us to explore the full space of mass deformations.
As we will see, the complete space of deformations of the $E_1^{(N)}$ theory is more involved, and exhibits a number of different phases separated by critical lines. The number of phases appears to grow with $N$, and the phases are qualitatively different for even and odd values of $N$. 
In the rest of this section we will analyze in detail the theories with $2\leq N\leq 6$, 
and then make some general observations about higher ranks.

\subsection{rank 2}

\begin{figure}[h!]
\center
\includegraphics[height=0.5\textwidth]{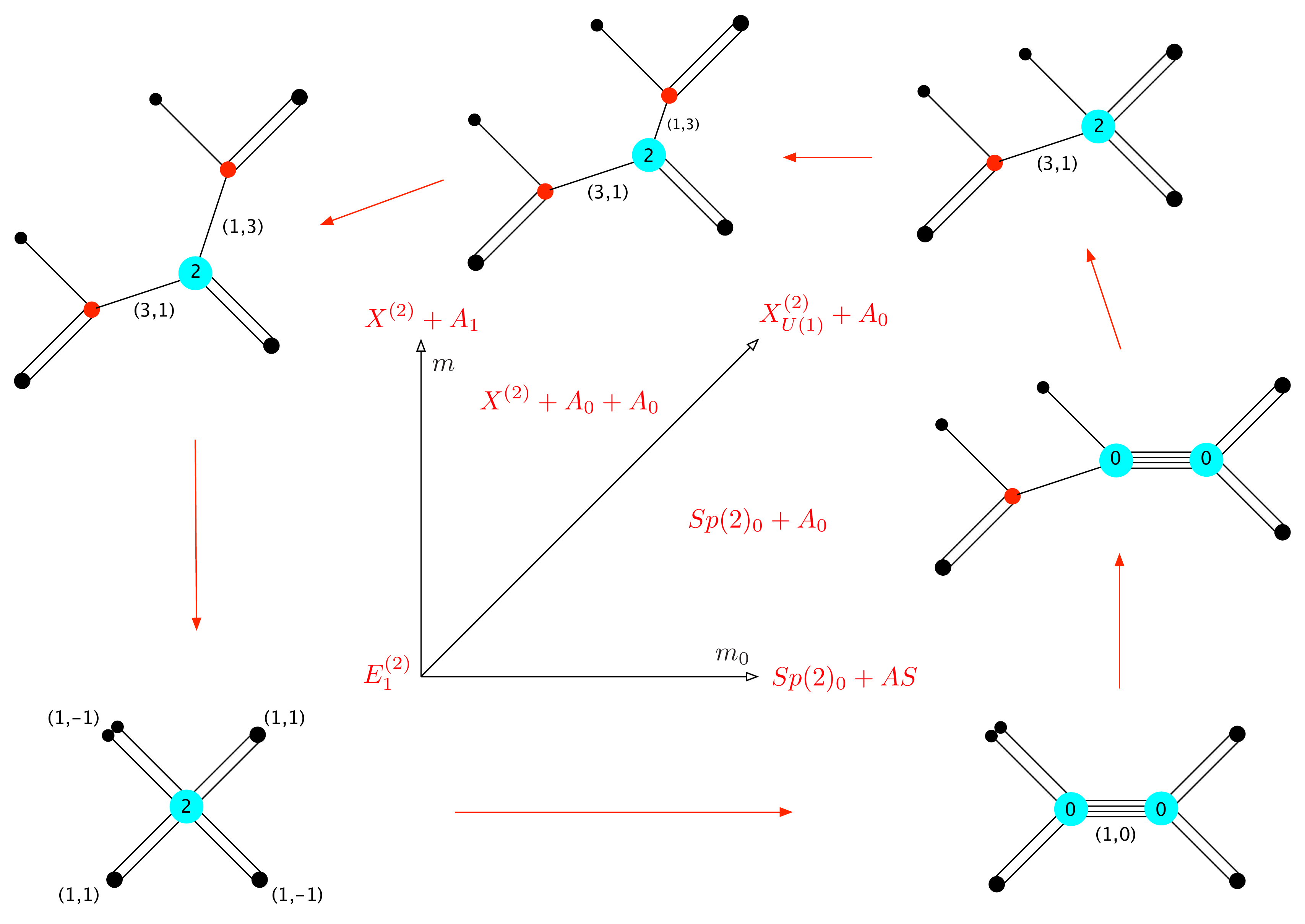} 
\caption{The phases of $E_1^{(2)}$. The number inside a non-trivial junction denotes its number
of local deformarions, {\em i.e.} Coulomb moduli.}
\label{E12Phases}
\end{figure}

The phase diagram of the rank 2 $E_1$ theory, and the corresponding series of 5-brane webs, are shown in Fig.~\ref{E12Phases}.\footnote{The 
phases of the rank 2 theory were originally identified using geometric engineering in \cite{Jefferson:2018irk}.}
We begin with the 5-brane junction for the $E_1^{(2)}$ theory in the lower left corner.
Moving in the $m_0$ direction 
leads to the 5-brane web in the lower right corner that describes the $Sp(2)_0 + AS$ theory \cite{Bergman:2015dpa}. 
The next deformation, corresponding to turning on $m$, leads to a 5-brane web 
containing a trivial junction.
The non-trivial part of this web describes the pure $Sp(2)_0$ theory \cite{Bergman:2015dpa}.
The trivial junction, which is equivalent to a detached (1,1) 7-brane, corresponds
to the massive antisymmetric hypermultiplet, namely to the $A_0$ part.
The next deformation corresponds to sending the $Sp(2)$ coupling to infinity, and leads to an interacting SCFT
that we denote by $X^{(2)}_{U(1)}$, meaning that it has rank 2 and a $U(1)$ global symmetry, and correspondingly a one-dimensional space of mass deformations.
This is the theory along the $m_0=m$ critical line.
Its deformation
in one direction gives the pure $Sp(2)_0$ theory.
The deformation in the other direction is shown in the next step, 
and describes an interacting theory with no
global symmetry. 
We therefore denote it generically as $X^{(2)}$.
There are also two identical 
trivial junctions in this case, and correspondingly two $A_0$ factors.
As we continue to deform in this direction we will reach a point where two equally charged states become massless
at the same point on the Coulomb branch. At this point the two $A_0$ factors enhance to $A_1$, 
and we get an $SU(2)$ global symmetry. 
In the web this corresponds to the $(3,1)$ and $(1,3)$ 5-branes having the same length.
This is the theory along the $m$ direction.
Finally, to complete the circuit we shrink the $(3,1)$ and $(1,3)$ 5-branes, 
and arrive back at the 5-brane junction for $E_1^{(2)}$.

\subsection{rank 3}

\begin{figure}[h!]
\center
\includegraphics[height=0.5\textwidth]{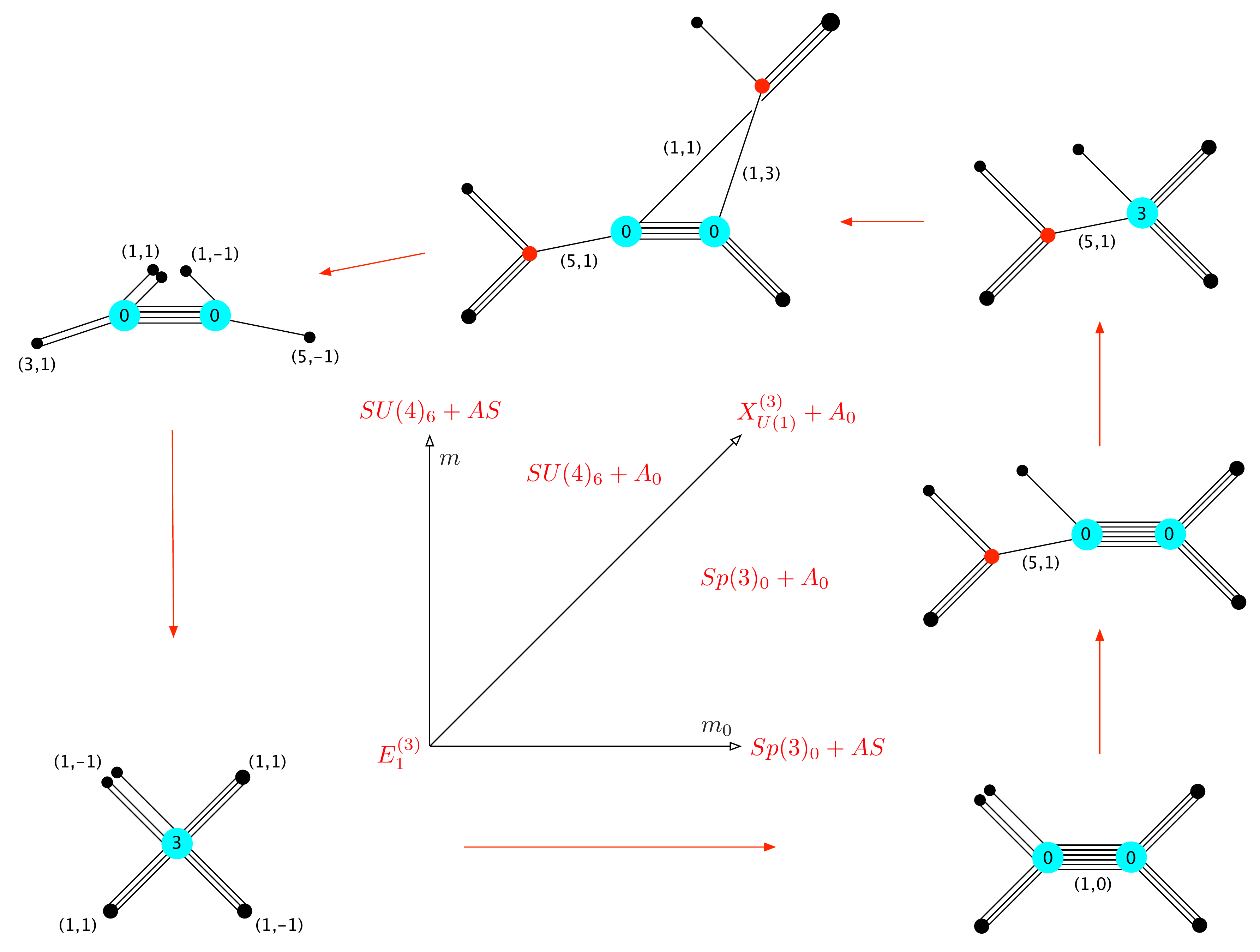} 
\caption{Phases of $E_1^{(3)}$}
\label{E13Phases}
\end{figure}

\begin{figure}[h!]
\center
\includegraphics[width=1\textwidth]{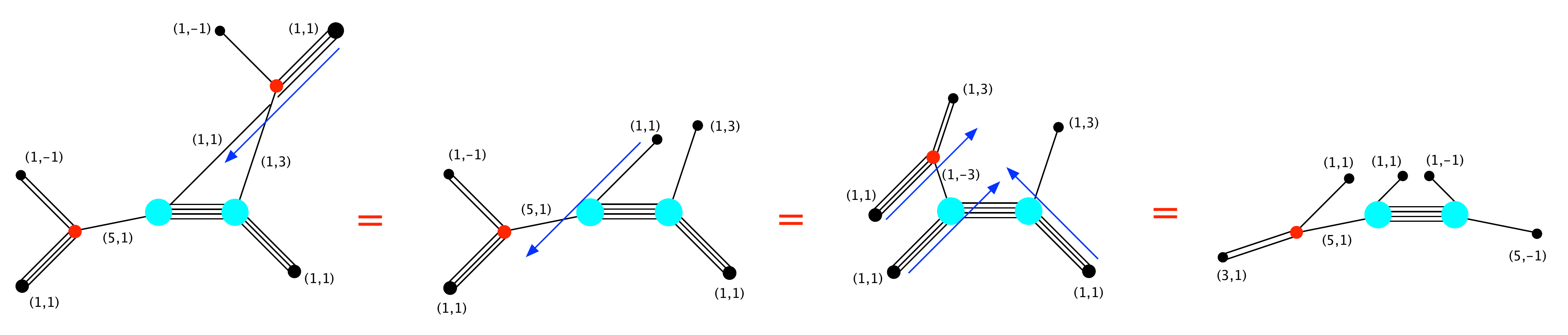} 
\caption{Equivalent 5-brane webs for $SU(4)_6$ with a massive antisymmetric hypermutiplet.}
\label{E13WebReduction}
\end{figure}

The phase diagram and corresponding 5-brane webs for the $E_1^{(3)}$ theory are shown in Fig.~\ref{E13Phases}.
The first three steps are very similar to the rank 2 case, ending at the 5-brane web describing a rank 3
interacting SCFT with a $U(1)$ global symmetry $X^{(3)}_{U(1)}$, 
together with a decoupled $A_0$ sector. 
This is the theory along the $m_0=2m$ critical line.
Deforming it in one direction gives the pure $Sp(3)_0$ theory.
However deforming it in the opposite direction gives something different from the previous case.
The complicated looking web resulting from this deformation describes 
an $SU(4)$ gauge theory with a CS level 6
with a massive hypermultiplet in the antisymmetric representation of $SU(4)$, namely the ${\bf 6}$ of 
$SU(4)=Spin(6)$.
This is seen by performing a number of Hanany-Witten moves that lead to the equivalent webs
shown in Fig.~\ref{E13WebReduction}.
The non-trivial part of the last 5-brane web in Fig.~\ref{E13WebReduction} describes $SU(4)_6$. 
Taking the mass corresponding to the trivial part to zero then leads
to the web in the upper left corner of Fig.~\ref{E13Phases}, which is a known representation of the theory $SU(4)_6 + AS$
\cite{Bergman:2015dpa}.
Since the rank 2 antisymmetric representation of $SU(4)$ is real, this 
theory enjoys an $SU(2)$ global symmetry, and must therefore appear along the $m$ direction.
The mass of the antisymmetric hypermultiplet is given by $m_0$.
Indeed deforming in either direction leads to $SU(4)_6$, since the shift in the CS level of an $SU(N)$ theory
due to a massive antisymmetric hypermultiplet is given by $\frac{1}{2}\mbox{sign}(m_0)(N-4)$, which vanishes for $N=4$.
Finally, taking $m\rightarrow 0$ corresponds to taking the $SU(4)$ coupling to infinity, and leads, 
using one of the other equivalent webs in Fig.~\ref{E13WebReduction}
and an $SL(2,\mathbb{Z})$ transformation, to
the original $E_1^{(3)}$ junction.
This implies, in particular, that the topological $U(1)$ symmetry of the $SU(4)_6 + AS$ theory is
enhanced, by instantons, to $SU(2)$ in the UV.

\subsection{rank 4}

\begin{figure}[h!]
\center
\includegraphics[height=0.5\textwidth]{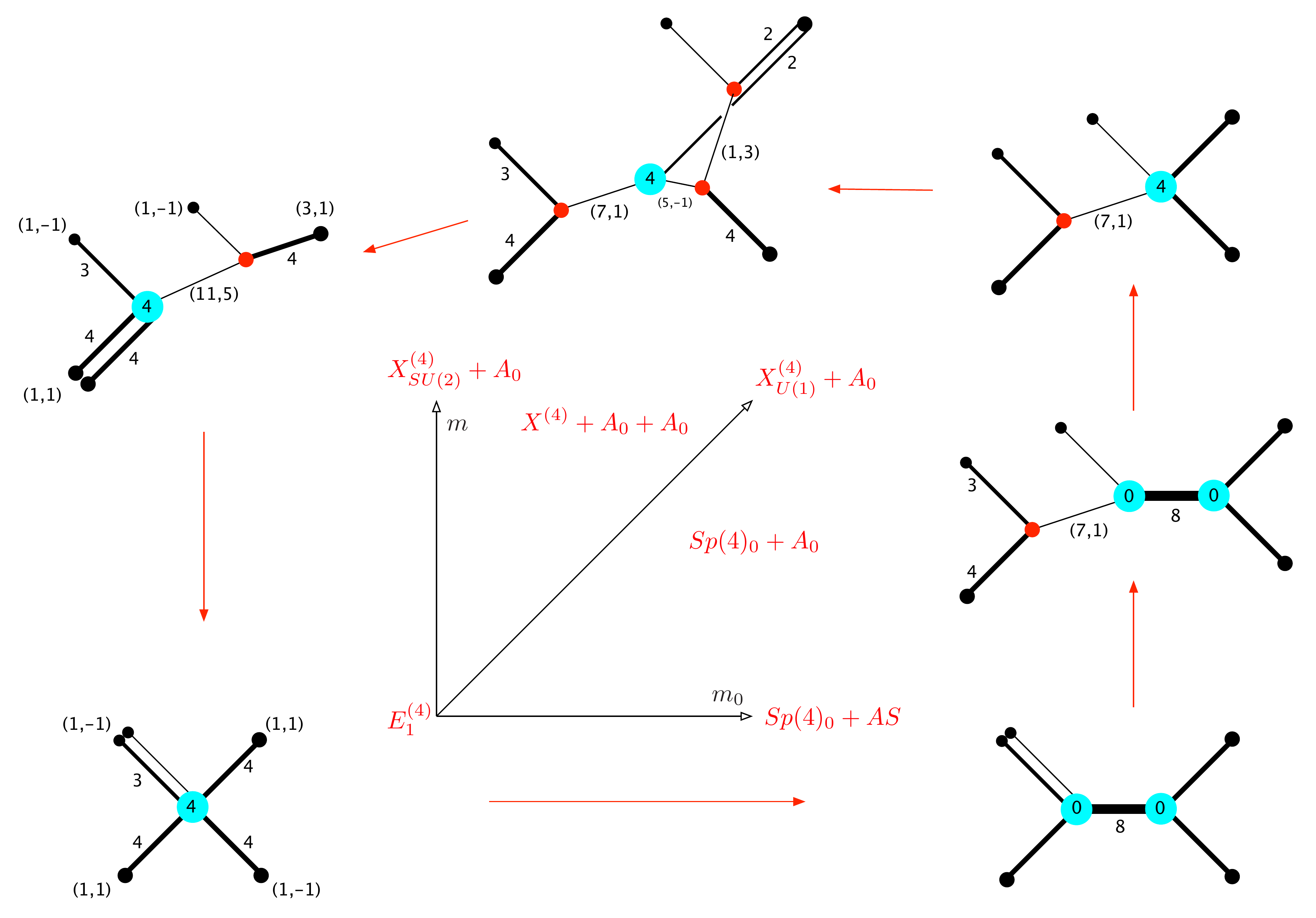} 
\caption{Phases of $E_1^{(4)}$}
\label{E14Phases}
\end{figure}

\begin{figure}[h!]
\center
\includegraphics[width=0.8\textwidth]{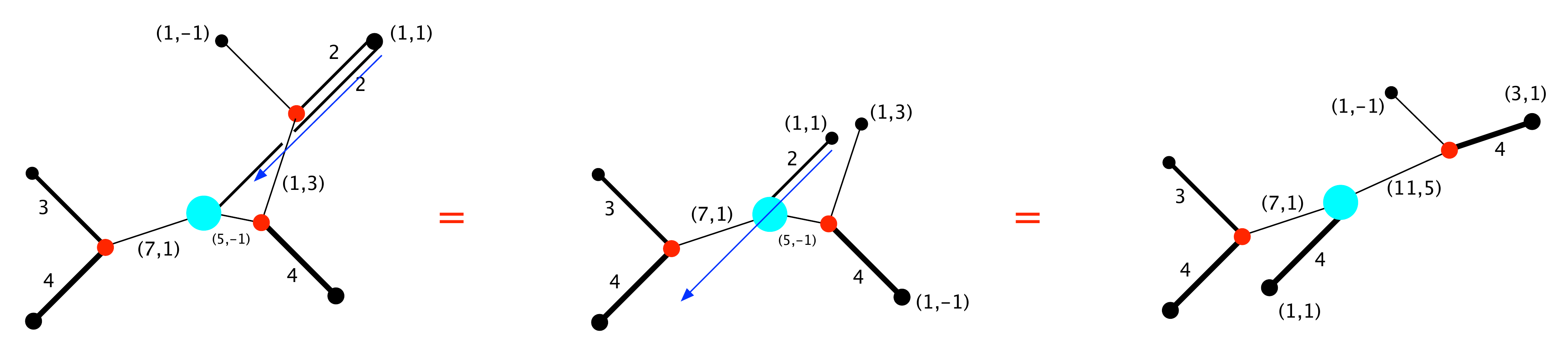} 
\caption{Equivalent webs for $X^{(4)}+ A_0 + A_0$.}
\label{E14WebReductions}
\end{figure}

The phase diagram and corresponding 5-brane webs for the $E_1^{(4)}$ theory are shown in Fig.~\ref{E14Phases}.
The first four steps are qualitatively similar to the rank 2 case, ending at at a 5-brane web describing
an interacting theory with no global symmetry $X^{(4)}$ together with two massive hypermultiplets. 
Equivalent webs for this theory are shown in Fig.~\ref{E14WebReductions}.
Unlike the rank 2 case, in this case the two massive hypermultiplets described by the two trivial 
junctions are different. They are equivalent to two detached 7-branes with different $(p,q)$ charges.
In the preceding deformation we turned on a mass for one of these hypermultiplets.
In the subsequent deformation we send the mass of the other one to zero, leading to the 5-brane web
in the upper-left corner of Fig.~\ref{E14Phases}. This describes a rank 4 interacting SCFT with an $SU(2)$ global symmetry,
realized by the pair of identical legs. Finally sending the mass of the remaining massive hypermultiplet to
zero leads back to (an $SL(2,\mathbb{Z})$ transform of) the $E_1^{(4)}$ junction.

\subsection{rank 5}

\begin{figure}[h!]
\center
\includegraphics[height=0.5\textwidth]{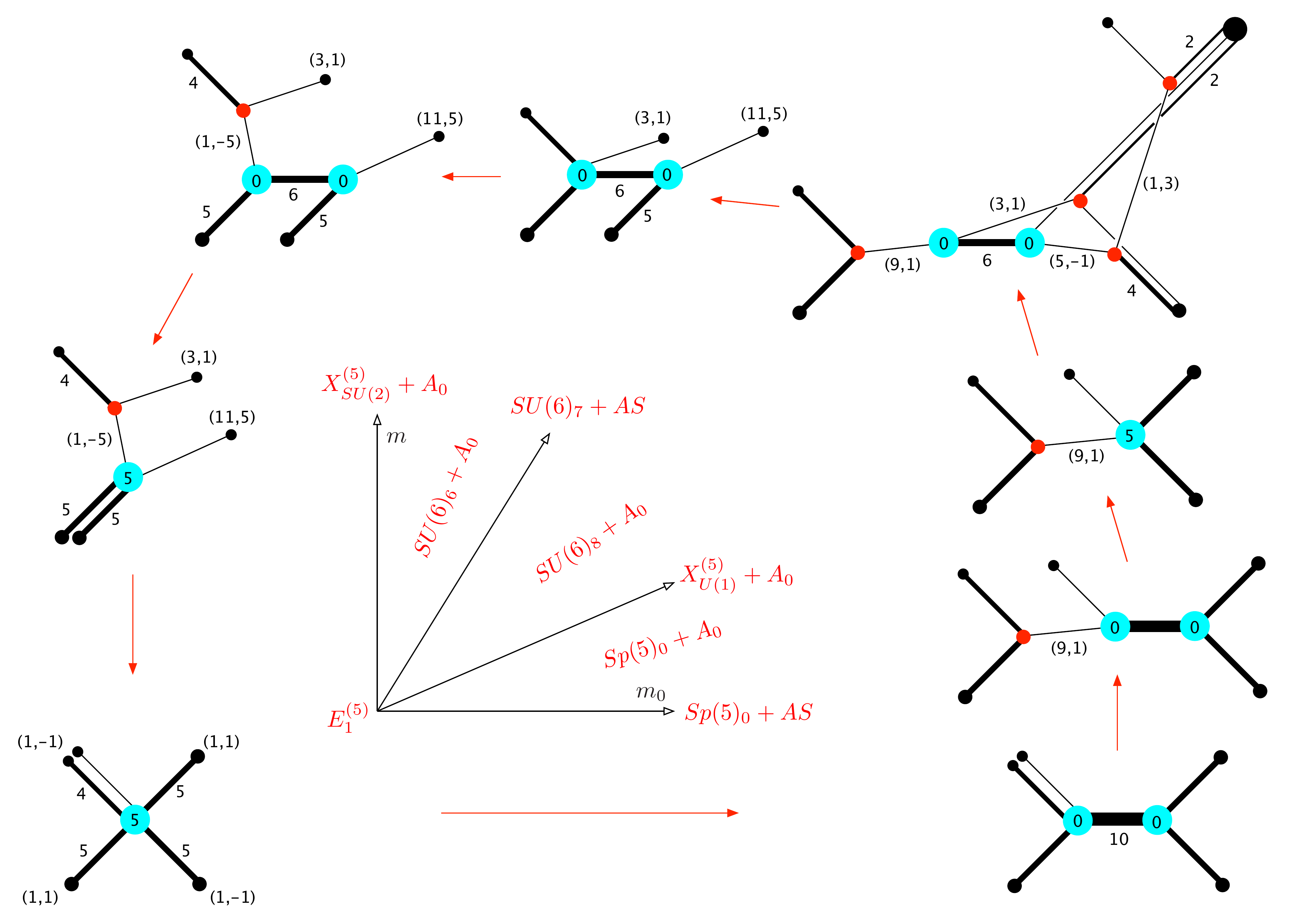} 
\caption{Phases of $E_1^{(5)}$}
\label{E15Phases}
\end{figure}

\begin{figure}[h!]
\center
\includegraphics[width=0.8\textwidth]{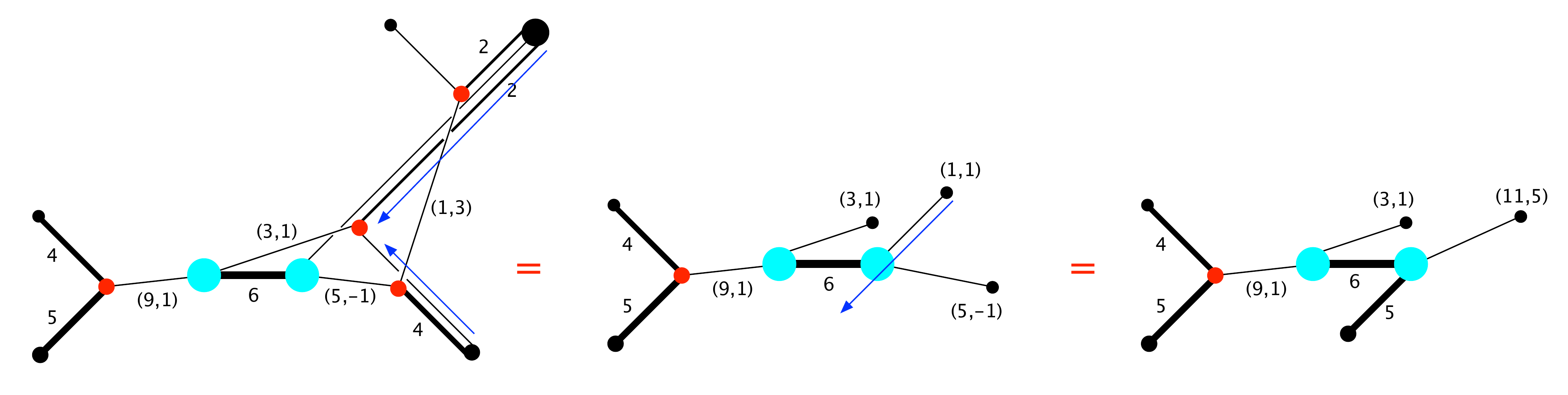} 
\caption{Equivalent webs for $SU(6)_8$ with a massive antisymmetric hypermultiplet.}
\label{E15WebReductions1}
\end{figure}

\begin{figure}[h!]
\center
\includegraphics[width=0.65\textwidth]{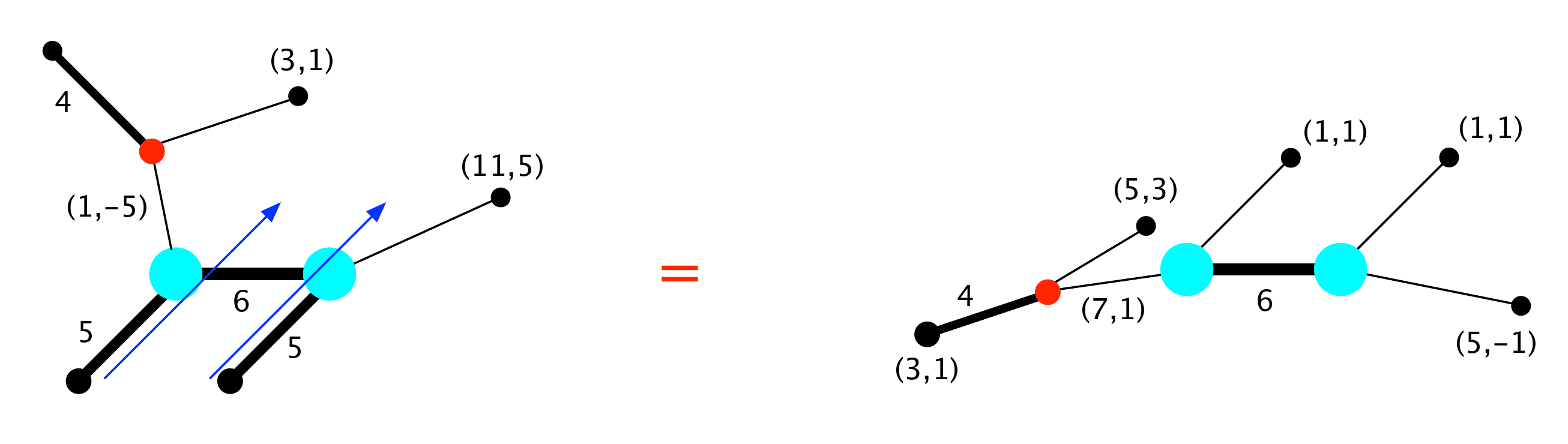} 
\caption{Equivalent webs for $SU(6)_6$ with a massive antisymmetric hypermultiplet.}
\label{E15WebReductions2}
\end{figure}

The phase diagram and 5-brane webs for the rank 5 theory $E_1^{(5)}$ are shown in Fig.~\ref{E15Phases}.
There are now three different phases separated by two critical lines.
The first few steps are analogous to the rank 3 case.
In particular the complicated looking web in the upper right corner describes an
$SU(6)_8$ theory with a massive antisymmetric hypermultiplet.
This can be seen in one of the equivalent webs shown in Fig.~\ref{E15WebReductions1}.
Taking the mass of the hypermultiplet to zero leads to the next web, which describes 
the $SU(6)_7 + AS$ theory.
Unlike the rank 3 case, the symmetry associated with the hypermultiplet is only $U(1)$, and so this theory 
lies on another diagonal critical line. 
A positive hypermultiplet mass shifts the CS level to $7 + \frac{6-4}{2} = 8$.
A negative mass shifts the CS level to $7 - \frac{6-4}{2} = 6$, which is the theory described by the next 
web in the upper left corner of Fig.~\ref{E15Phases}.
This is seen in the equivalent web in Fig.~\ref{E15WebReductions2}.
In the next deformation we take the $SU(6)_6$ coupling to infinity, which leads to an interacting
SCFT with an $SU(2)$ global symmetry. 
We see this explicitly in the next 5-brane web, but this is also more
generally a property of the $SU(N)_N$ theory \cite{Bergman:2013aca}.
Closing the circuit back to the $E_1^{(5)}$ junction then just follows by sending the remaining 
mass to zero
and performing a Hanany-Witten move.
We end this subsection by noting that 
the $U(1)\times U(1)$ global symmetry of the 
$SU(6)_7 + AS$ theory must be enhanced by instantons to $SU(2)\times SU(2)$ in the UV, since this corresponds to sliding back to the $E_1^{(5)}$ point at the origin along the critical line.

\subsection{rank 6}

\begin{figure}[h!]
\center
\includegraphics[height=0.5\textwidth]{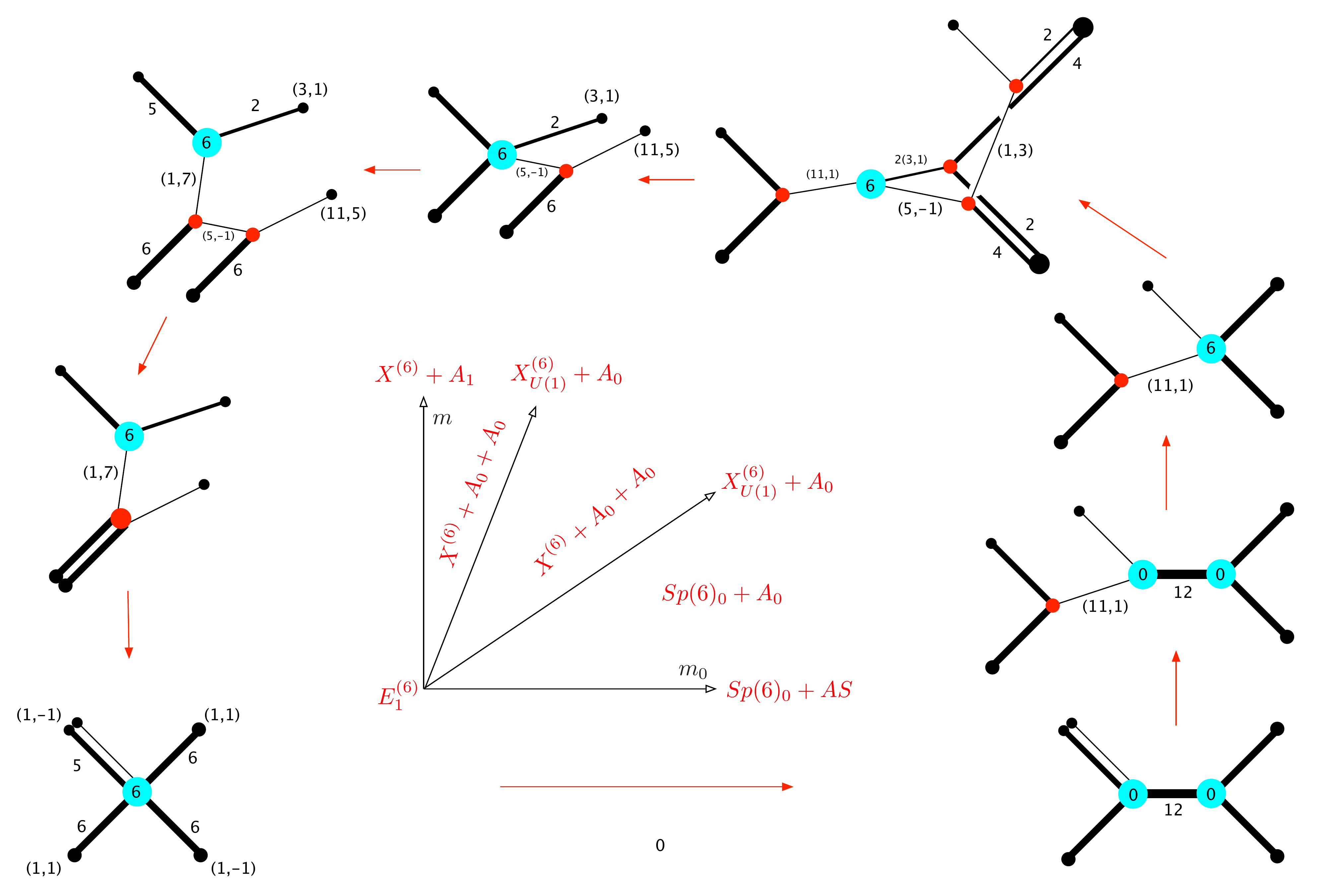} 
\caption{Phases of $E_1^{(6)}$}
\label{E16Phases}
\end{figure}

The phase diagram and 5-brane webs for the rank 6 theory $E_1^{(6)}$ are shown in Fig.~\ref{E16Phases}.
There are three different phases separated by two critical lines.
The steps from $E_1^{(6)}$ to $X^{(6)} + A_0 + A_0$ are similar to the rank 4 case.
At the next step we encounter a 5-brane web that describes another rank 6 interacting theory with a $U(1)$ global symmetry, which is distinct from the one corresponding to the infinite coupling limit of the $Sp(6)_0$ theory
(although we use the same generic notation $X^{(6)}_{U(1)}$).
Continuing to deform in the same direction leads to a web describing another rank 6 interacting theory with no global symmetry. 
The next deformation merges the two $A_0$ sectors to an $A_1$ sector, {\em i.e.} two identical massive hypermultiplets.
This is the theory along the $m$ axis.
The final deformation sends the mass to zero, and recovers the $E_1^{(6)}$ configuration after a
Hanany-Witten move.

\subsection{rank N}

The number of different phases of the $E_1$ theory appears to grow with increasing rank $N$,
and the pattern is qualitatively different for odd and even ranks.
The latter observation is consistent with the fact that $E_1$ theories whose ranks differ
by an even number can be connected by going on the Coulomb branch,
but $E_1$ theories with an odd difference of ranks cannot.
We can see this explicitly in the initial gauge theory deformation.
Start with the $Sp(N)_0 + AS$ theory and go to a point on the Coulomb branch 
where $Sp(N) \rightarrow Sp(N-1) \times U(1)$.
The $Sp(N)$ antisymmetric decomposes into a singlet, a neutral $Sp(N-1)$ antisymmetric, 
and a charged $Sp(N-1)$ fundamental and its conjugate.
The latter two states are massive, and their masses have opposite signs.
Integrating these states out therefore shifts the mod 2 theta parameter 
of the $Sp(N-1)$ factor relative to the original $Sp(N)$ theory.
Going down further to $Sp(N-2)$ shifts it back to its original value, etc.
Indeed we will see in the next section that the deformation pattern of the $\tilde{E}_1$
theory, whose initial gauge theory deformation gives $Sp(N)_\pi + AS$,
will similarly alternate between even and odd ranks.

\begin{figure}[h!]
\center
\includegraphics[width=0.7\textwidth]{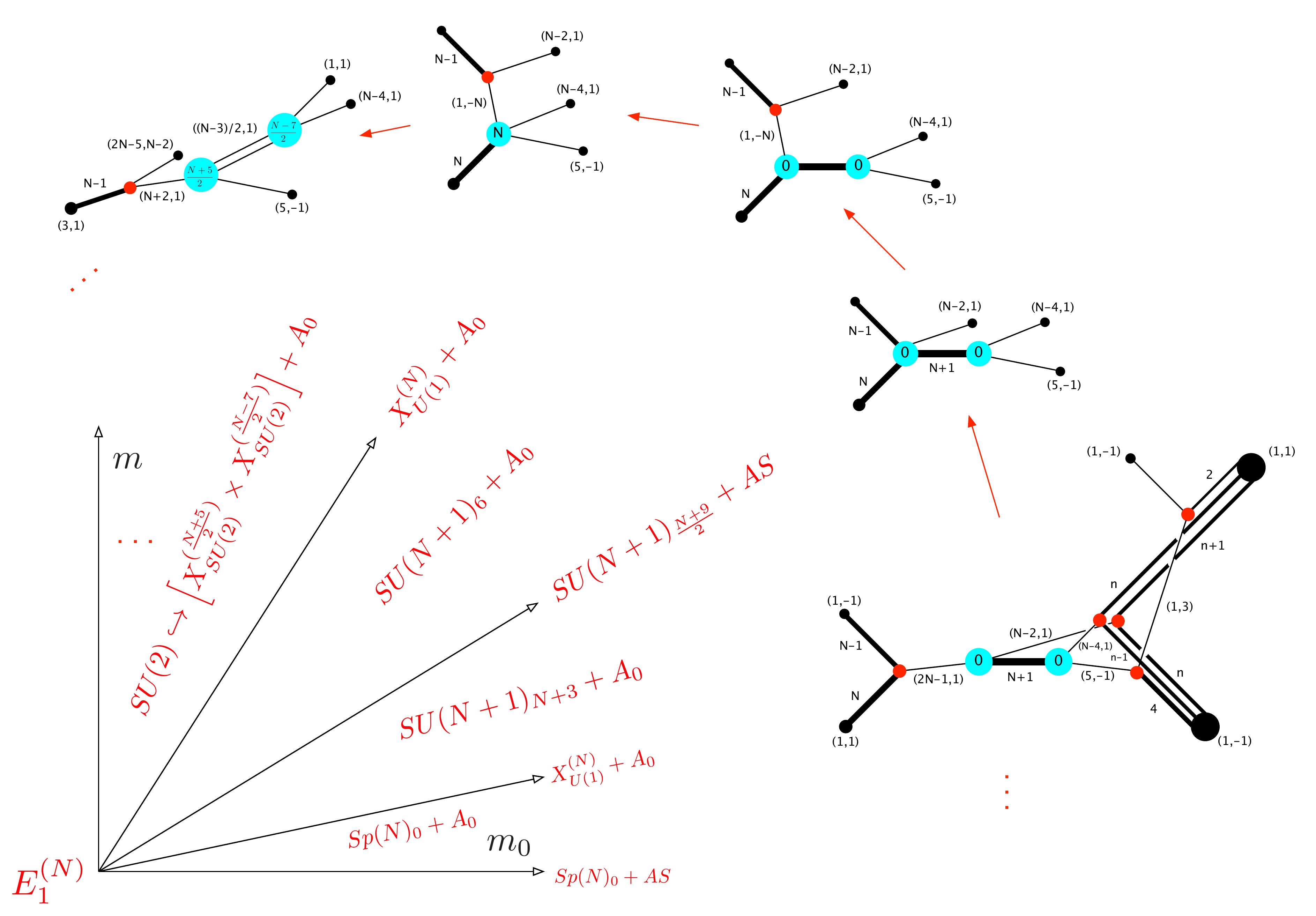} 
\caption{The first few phases of $E_1^N$ for odd $N\geq 7$.}
\label{E1NOddPhases}
\end{figure}

Though we have not been able to map out the complete phase diagram for $N>6$,
we can offer some interesting observations about the first few phases,
especially for odd $N$, Fig.~\ref{E1NOddPhases}.
The even rank case is less illuminating, so we omit it. 
The first critical line separates the phase described by $Sp(N)_0$ and $SU(N+1)_{N+3}$.
This ``duality" was first studied in \cite{Gaiotto:2015una}.
Moving counter-clockwise, the next critical line we encounter corresponds to the theory
$SU(N+1)_{\frac{N+9}{2}} + AS$. A positive hypermultiplet mass leads to $SU(N+1)_{N+3}$,
and a negative hypermultiplet mass to $SU(N+1)_6$.
We are also lead to the general prediction that the $U(1)\times U(1)$ global symmetry of
$SU(N+1)_{\frac{N+9}{2}} + AS$ for odd $N$ is enhanced in the UV to $SU(2)\times SU(2)$.
The next critical line corresponds to the infinite coupling limit of the $SU(N+1)_6$ theory, which for
$N>5$ does not have an enhanced $SU(2)$ symmetry. 
Beyond this critical line the web deforms as shown in the last step in Fig.~\ref{E1NOddPhases}.
This describes a product of two interacting SCFT's, each of which has an $SU(2)$ global symmetry, with the diagonal $SU(2)$ gauged.
We denote this as $SU(2)\hookrightarrow\left[X_{SU(2)}^{(\frac{N+5}{2})}\times X_{SU(2)}^{(\frac{N-7}{2})}\right]$. 
In fact these SCFT's correspond separately to the UV fixed points of 
$SU(\frac{N+7}{2})_{\frac{N+7}{2}}$ and $SU(\frac{N-5}{2})_{\frac{N-5}{2}}$.\footnote{This generalizes a recent observation made in
\cite{Minahan:2020ifb}, that the ``continuation past infinite coupling" of $SU(N)_0$ is given by 
$SU(\frac{N}{2})_{\frac{N}{2}} \times SU(\frac{N}{2})_{-\frac{N}{2}}\times SU(2)$.}
This is as far as we got.
The next critical line will correspond to taking the mass of the hypermultiplet to zero, and so on.

\section{The $\tilde{E}_1$ theory}

The $\tilde{E}_1$ theory is a variant of the $E_1$ theory.
At rank 1 this is a theory with a $U(1)$ global symmetry, that with a positive mass deformation flows in the IR to the
supersymmetric $SU(2)$ gauge theory with a non-trivial theta parameter $\theta=\pi$.
Since the global symmetry is only $U(1)$, the negative mass deformation may lead to a different theory,
and indeed the IR theory is an interacting SCFT without any global symmetry known as the $E_0$ theory.
In the Type IIA string theory construction, the $\tilde{E}_1$ theory is realized by turning on an additional torsion-valued RR flux \cite{Bergman:2013ala}. 
Just as in the previous case, we can consider the rank $N$ generalization of this set-up corresponding to $N$ D4-branes, denoted by $\tilde{E}_1^{(N)}$. 
This theory has a $U(1)\times SU(2)$ global symmetry and a two-dimensional space of supersymmetric mass parameters $(m_0,m)$.
In this case there is a reflection symmetry only along the $m$ direction, so we will need to consider the entire
upper half plane.
Along the positive $m_0$ direction, the theory flows to the $Sp(N)_\pi +AS$ theory, which has a topological $U(1)$ symmetry and a matter $SU(2)$ symmetry. Along the negative $m_0$ direction, we expect it to flow to a rank $N$ generalization of the $E_0$ theory with an $SU(2)$ global symmetry which we will denote by $E_0^{(N)}$.
The same perturbative analysis as before shows that there exists a critical line $m_0=(N-1)m$ corresponding to the infinite coupling limit of the pure supersymmetric $Sp(N)_\pi$ gauge theory.
As before, we will determine what happens beyond this line using 5-brane webs.
The pattern that will emerge for $\tilde{E}_1$ with even rank will resemble the pattern for $E_1$ with odd rank,
and vice versa.
We will therefore be somewhat briefer in our exposition, presenting explicitly 
only the rank 2,3 and 4 cases, and we will also refrain from presenting 
the equivalent webs that were useful in identifying some of the phases.

\subsection{rank 2}

\begin{figure}[h!]
\center
\includegraphics[height=0.6\textwidth]{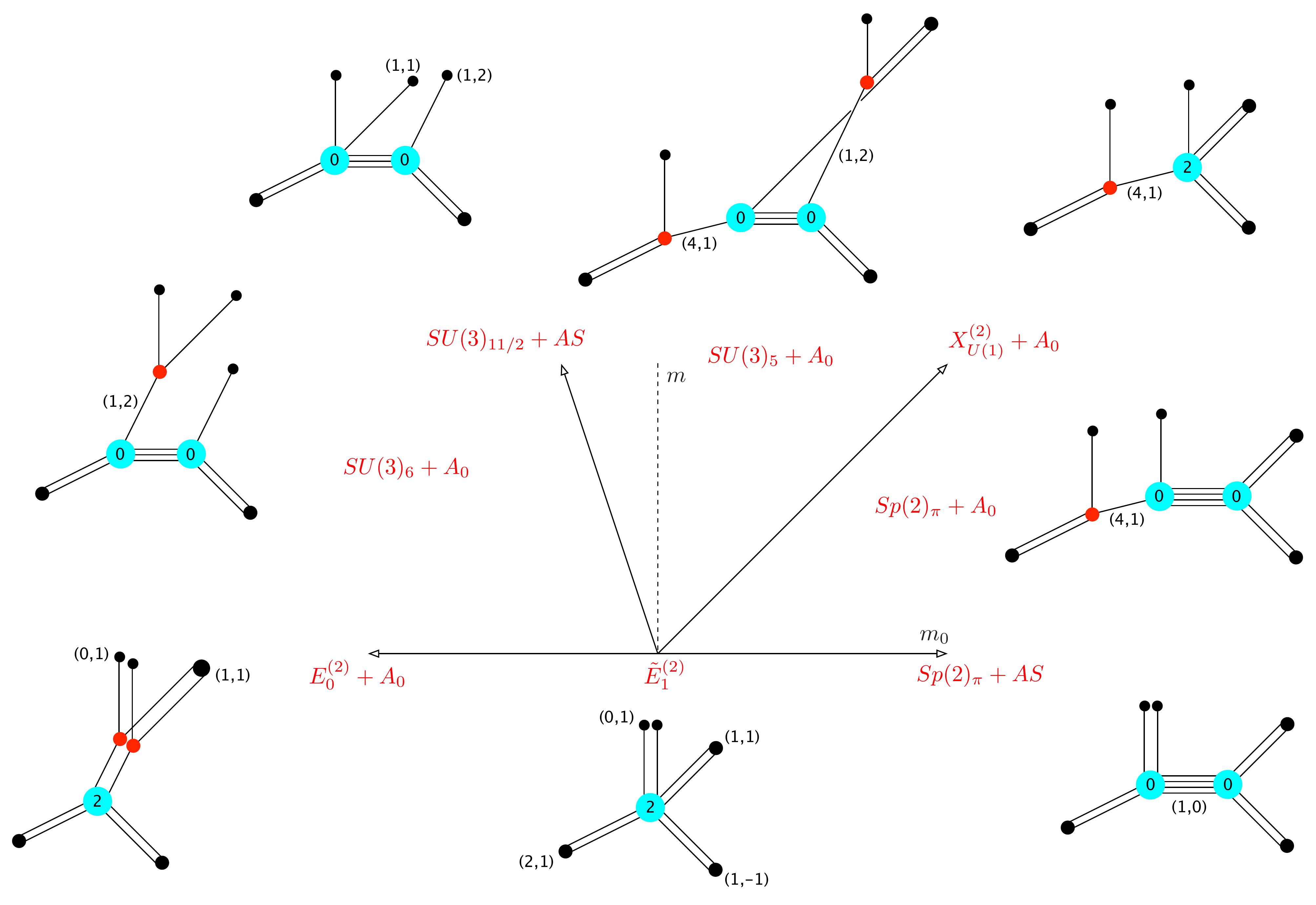} 
\caption{Phases of $\tilde{E}_1^{(2)}$}
\label{TildeE12Phases}
\end{figure}

The phase diagram for $\tilde{E}_1^{(2)}$ and the corresponding 5-brane webs are shown in 
Fig.~\ref{TildeE12Phases}.\footnote{The 
phases of the rank 2 theory were originally identified using geometric engineering in \cite{Jefferson:2018irk}.}
Deforming along $m_0>0$ leads to a web that describes the $Sp(2)_\pi + AS$ theory \cite{Bergman:2015dpa}. 
Then deforming along $m$ leads to a web whose non-trivial part describes the pure $Sp(2)_\pi$
theory \cite{Bergman:2015dpa}.
The critical line corresponds to the infinite coupling limit of the $Sp(2)_\pi$ theory,
which is an interacting SCFT with a $U(1)$ global symmetry, $X^{(2)}_{U(1)}$.
The next deformation, the ``contiuation past infinite coupling" of
the $Sp(2)_\pi$ theory, leads
to a web that is easily shown, by a couple of equivalence moves, to describe the gauge theory
$SU(3)_5$ with a massive fundamental (or equivalently antisymmetric) hypermultiplet.
Taking the mass of the hypermultiplet to zero leads to the web in the upper left corner,
which corresponds to the second critical line. On this line the theory flows to $SU(3)_{11/2} + AS$. 
This also leads to the prediction that the $U(1)\times U(1)$ global symmetry 
of the $SU(3)_{11/2} + AS$ theory is enhanced at infinite coupling to $SU(2)\times U(1)$.
Turning on the opposite sign mass for the hypermultiplet then
gives $SU(3)_6$, corresponding to the next web in the figure.
The next deformation takes the gauge coupling to infinity, and leads to the theory described by 
the web in the lower left corner, which is the rank 2 version of the $E_0$ theory. 
Finally, we take the mass of the hypermultiplet to zero,
and recover the $\tilde{E}_1^{(2)}$ junction.

\subsection{rank 3}

\begin{figure}[h!]
\center
\includegraphics[height=0.6\textwidth]{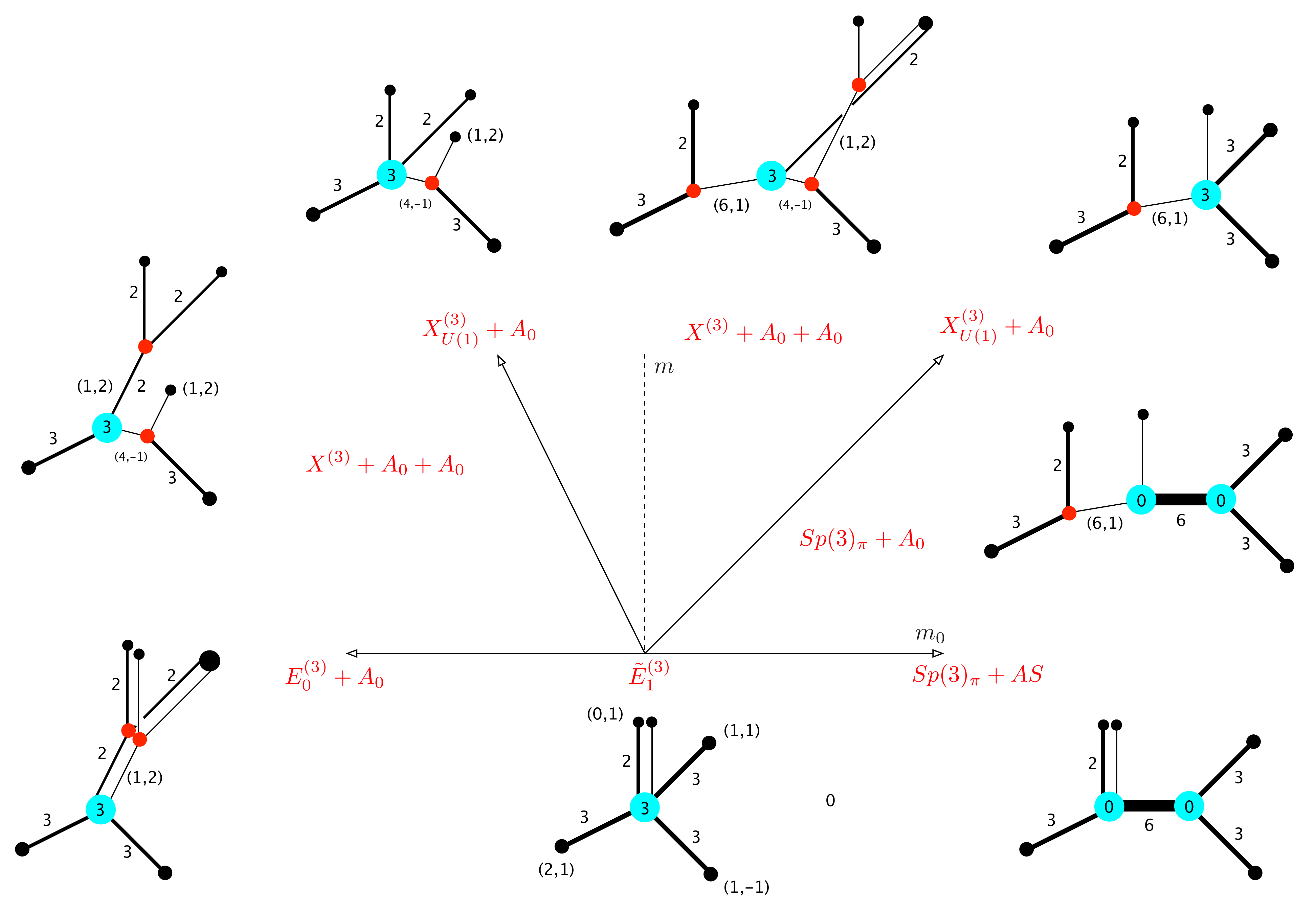} 
\caption{Phases of $\tilde{E}_1^{(3)}$}
\label{TildeE13Phases}
\end{figure}

The phase diagram and 5-brane webs for $\tilde{E}_1^{(3)}$ are shown in 
Fig.~\ref{TildeE13Phases}.
Other than the $Sp(3)_\pi$ phase, all phases are interacting theories.

\subsection{rank 4}

\begin{figure}[h!]
\center
\includegraphics[height=0.6\textwidth]{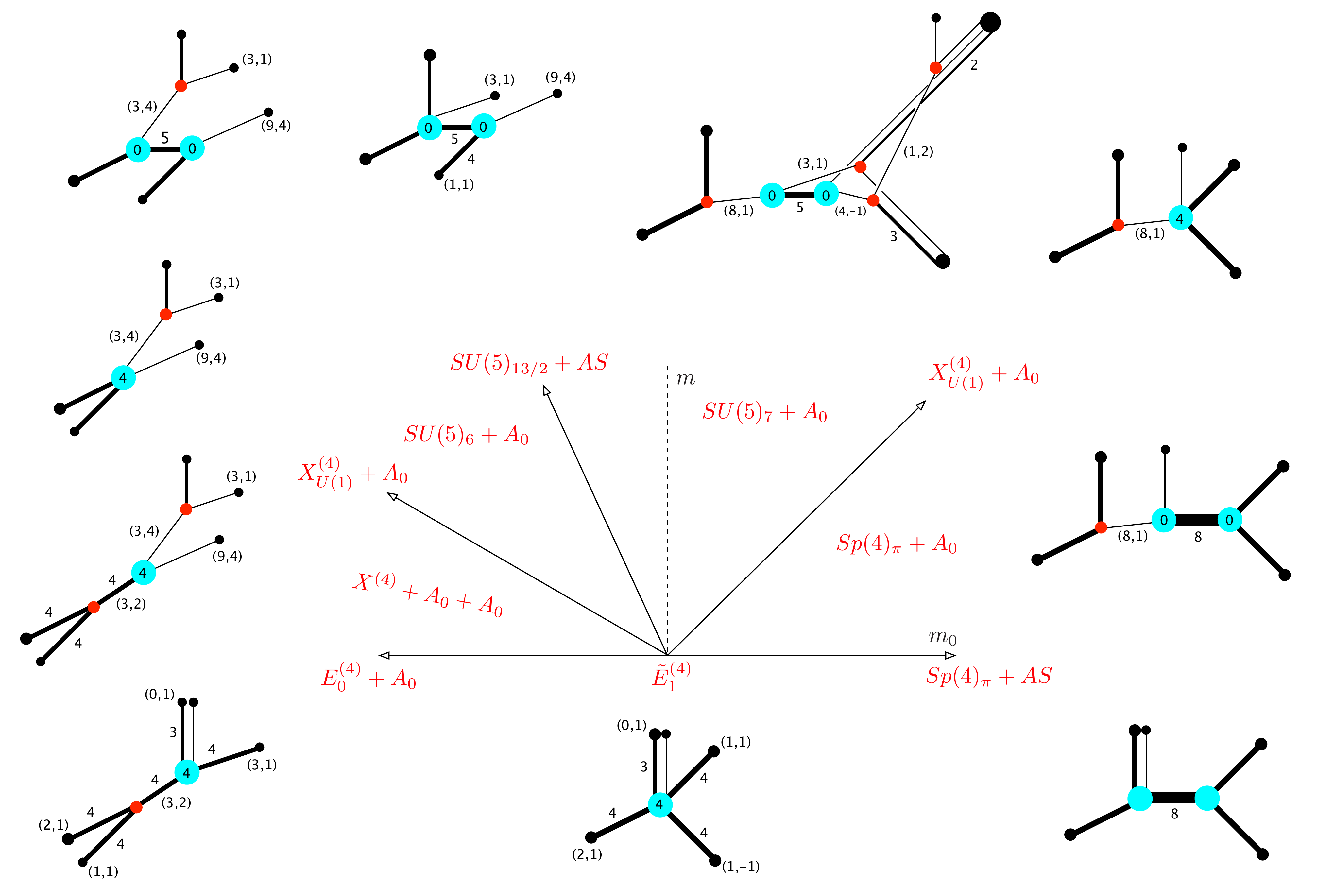} 
\caption{Phases of $\tilde{E}_1^{(4)}$}
\label{TildeE14Phases}
\end{figure}

The phase diagram and 5-brane webs for $\tilde{E}_1^{(4)}$ are shown in 
Fig.~\ref{TildeE14Phases}.
There is an additional critical line and an additional phase compared to $\tilde{E}_1^{(2)}$.
As in the rank 2 case, we predict that the $U(1)\times U(1)$ symmetry
of the $SU(5)_{13/2} + AS$ theory is enhanced at infinite coupling to $SU(2)\times U(1)$.

\subsection{rank $N$}

\begin{figure}[h!]
\center
\includegraphics[width=0.7\textwidth]{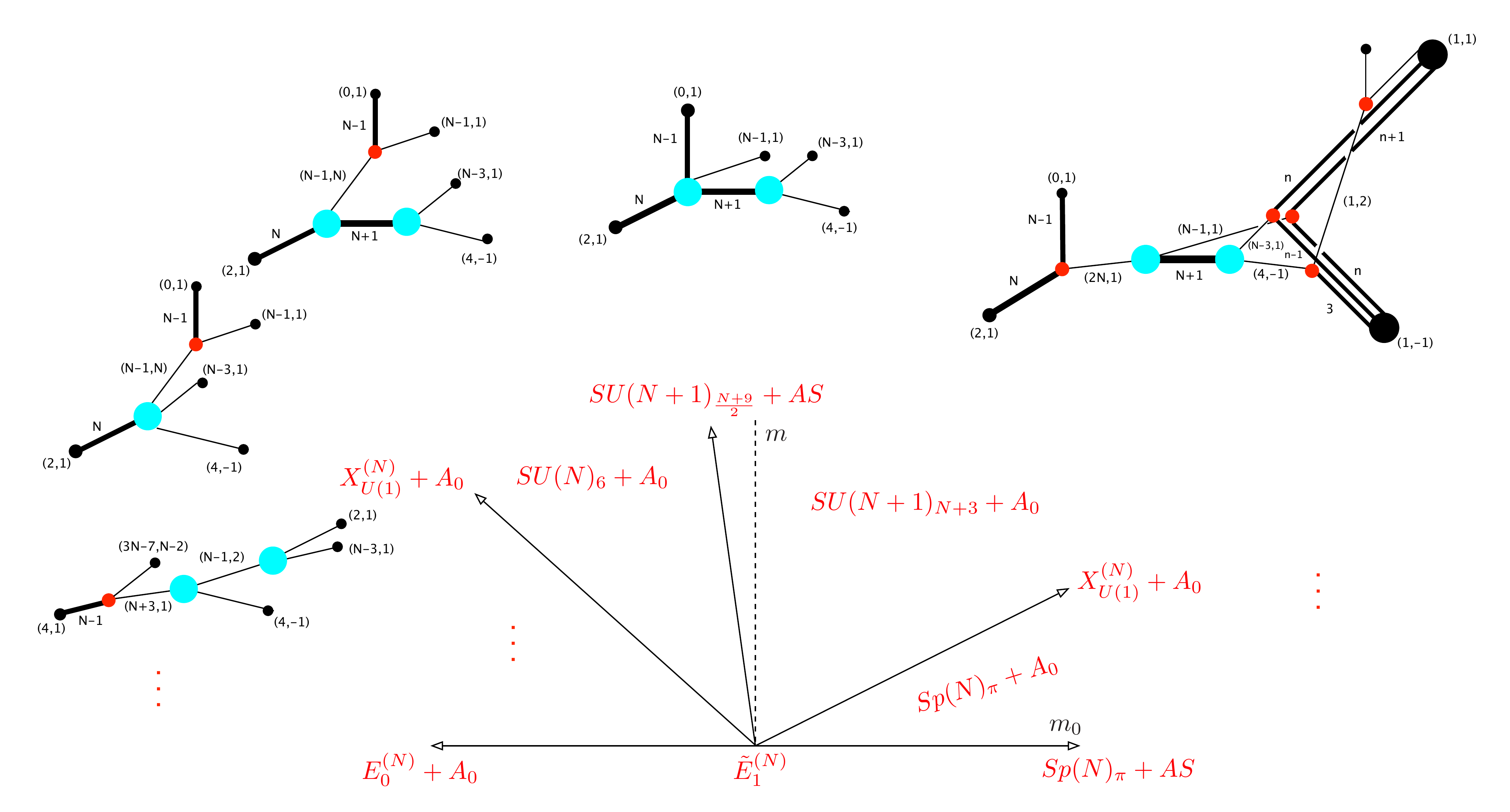} 
\caption{The first few phases of $\tilde{E}_1^N$ for even $N\geq 6$.}
\label{TildeE1NEvenPhases}
\end{figure}

As for the $E_1$ theory, the number of phases appears to grow with increasing $N$,
and the pattern is different for odd and even ranks.
The IR theories arising in the deformations of the odd rank $\tilde{E}_1^{(N)}$ theories 
are of the same type as those arising in the deformations of the even rank $E_1^{(N)}$ theories,
and vice versa.
This is consistent with the fact that $E_1$ and $\tilde{E}_1$ theories whose rank differs by
an odd number can be related by Coulomb branch deformations.
In Fig.~\ref{TildeE1NEvenPhases} we show the first few phases of the $\tilde{E}_1^{(N)}$ 
theory with even $N\geq 6$.
The first critical line separates the $Sp(N)_\pi$ phase and the $SU(N+1)_{N+3}$ phase,
which is again in agreement with the duality studied in \cite{Gaiotto:2015una}.
The subsequent steps are 
similar to the odd rank $E_1^{(N)}$ theory.
Our last web in the lower left corner is somewhat interesting, as it describes a theory obtained by 
taking two interacting SCFT's that do not have any global symmetry, and ``coupling" them
with a ``bi-fundamental" hypermultiplet, and another massive hypermultplet described by the 
trivial junction.
The next critical line will correspond to taking the mass of the latter to zero, and so on.

\section{Conclusions}

In this paper we have analyzed the two-dimensional space of mass deformations of two families of 
5d superconformal field theories given by the higher rank generalizations of the $E_1$ and $\tilde{E}_1$ theories.
We have seen that this space contains a number of different IR phases 
that grows with the rank of the theory.
A number of these phases are described by IR free supersymmetric gauge theories, while others 
are interacting SCFT's.
One of these phases is the 5d $SU(N)$ theory with CS level $k=\frac{N}{2} + 4$ and an antisymmetric hypermultiplet, 
which is one of the more recent gauge theories argued to possess a UV fixed point \cite{Bhardwaj:2020gyu}.
Its appearance in the space of deformations of the odd rank $E_1$ theory and the even rank $\tilde{E}_1$ theory
implies that it must exhibit an enhanced global symmetry at strong coupling, 
$SU(2)\times SU(2)$ for even $N$ and $SU(2)\times U(1)$ for odd $N$. 
This enhancement has also recently been suggested on the basis 
of geometrical engineering/reduction from 6d in \cite{Bhardwaj:2020avz}. It would be interesting to 
confirm this by field-theoretic means, using for example the superconformal index.

Though we have not been able to fully map the space of deformations in the general case,
we have made the general observation that the space of deformations of the even rank $E_1$ theory is 
qualitatively similar to that of the odd rank $\tilde{E}_1$ theory, and vice versa.
This can also be understood from the fact that the even/odd rank $E_1$ theories are related to the
odd/even rank $\tilde{E}_1$ theories by going on the Coulomb branch.

Our main tool has been the construction of 5d SCFT's using 5-brane junctions in Type IIB string theory,
but we expect there to be a parallel description of this using geometric engineering in M-theory,
generalizing the analysis of the rank 2 theories in \cite{Jefferson:2018irk}.
Our preliminary investigation of the $E_1$ and $\tilde{E}_1$ theories indicate that higher rank 
SCFT's in five dimensions possess a zoo of different IR phases depending on the direction one
takes in the space of mass deformations. This clearly deserves further study.
In particular, it would be interesting to identify the massless BPS states associated to the critical lines 
separating the different phases.

Lastly, a particularly interesting aspect of the high rank $E_1$ and $\tilde{E}_1$ theories
is that they have a holographic $AdS_6$ dual in massive type IIA string theory \cite{Brandhuber:1999np}.  
It would be interesting find the holographic description of the mass deformations studied in this paper.

\section*{Acknowledgments}

We thank G.~Zafrir for useful discussions. The work of O.B.  is supported in part by the Israel Science Foundation under grant No. 1390/17. The work of D.R.G.~is supported in part by the Spanish government grant MINECO-16-FPA2015-63667-P, and by the Principado de Asturias through the grant FC-GRUPIN-IDI/2018/000174.

\appendix

\section{5-brane webs and their deformations} 

Type IIB string theory admits 5-branes charged magnetically under both the NSNS 2-form $B_2$ and the 
RR 2-form $C_2$. The minimally charged objects are $(p,q)$ 5-branes, where $p$ and $q$ are relatively prime 
integers labeling the RR and NSNS charge, respectively.
In particular the $(1,0)$ 5-brane is the D5-brane and the $(0,1)$ 5-brane is the NS5-brane.
A 5-brane web is a configuration of Type IIB $(p,q)$ 5-branes containing 5-brane segments and vertices,
or junctions (which are really 4d spaces), where three or more 5-branes meet.
The Gauss law requires the sum of charges to vanish at each junction,
$\sum_{i\in {\cal J}} p_i = \sum_{i\in {\cal J}} q_i = 0$.
Supersymmetry further restricts the entire configuration to be planar, and the
relative orientation of each $(p,q)$ 5-brane in this plane to be given by 
$p+q\tau$ (or $p + q\bar{\tau}$), 
where $\tau$ is the complex Type IIB axion-dilaton. For clarity of presentation one usually 
assumes that $\tau = i$. 
There are analogous configurations of $(p,q)$ strings.

A planar configuration consisting of $N\geq 3$ 5-branes meeting at a point, namely an $N$-junction,
describes a 
five dimensional superconformal field theory \cite{Aharony:1997ju,Aharony:1997bh}.
The parameters and moduli of the 5d theory are realized as geometric deformations of the
5-brane junction. In particular global deformations that move the external 5-branes correspond
to the mass parameters of the field theory, and local deformations that keep the external 5-branes
fixed correspond to the Coulomb moduli of the field theory.
The latter appear as faces in the resulting 5-brane web, and the dimension of the Coulomb branch
is given by the number of faces in the 5-brane web. 
The number of mass parameters, and therefore the rank of the global symmetry, is given by $N-3$. 
This is because moving the junction as a whole 
does not change the theory, and fixing the orientations of $N-1$ of the 5-branes
determines that of the remaining 5-brane.
In cases where the mass-deformed 5-brane web consists of parallel 5-brane segments 
the corresponding theory is a supersymmetric gauge theory, in which the inverse-squared-YM coupling
is given by the value of the mass.

\subsection{7-branes and equivalent webs} 

For a more complete description and classification of 5d SCFT's we need to introduce $(p,q)$ 7-branes
on which the $(p,q)$ 5-branes of the 5-brane junction end.
For a 7-brane the $(p,q)$ charges define an 
$SL(2,\mathbb{Z})$ monodromy action as one encircles it.
The monodromy occurs across a branch cut emanating from the 7-brane.
The clockwise monodromy around a $(p,q)$ 7-brane is given by
\be
M_{(p,q)} = \left(
\begin{array}{cc}
1-pq & p^2 \\
-q^2 & 1+pq
\end{array}
\right) \,.
\ee
In our convention the $(1,0)$ 7-brane is a D7-brane,
so that a $(p,q)$ 5-brane ends on a $(p,q)$ 7-brane.\footnote{Note that strictly speaking 5-branes follow geodesics in the background of the 7-branes. However, one may push the 7-branes far away from the meeting point where the SCFT lives by extending the external 5-branes, such that for all practical purposes, the axion-dilaton is approximatelty constant over the relevant scales of the field theory.}

The addition of the 7-brane endpoints introduces a number of additional deformations.
The first corresponds to the possibility of breaking the 5-brane junction into 5-brane segments or
sub-junctions that can move separately along the 7-branes, namely perpendicular to the plane of the junction.
This describes the Higgs branch moduli of the field theory.

The second corresponds to the motion of the 7-branes in the plane.
For each 7-brane we can divide this into a component perpendicular the 5-branes ending on it,
and a component parallel to them.
The former are just the same mass parameters that we previously identified with the global deformations 
that move the external 5-branes.

\begin{figure}[h!]
\center
\includegraphics[width=0.7\textwidth]{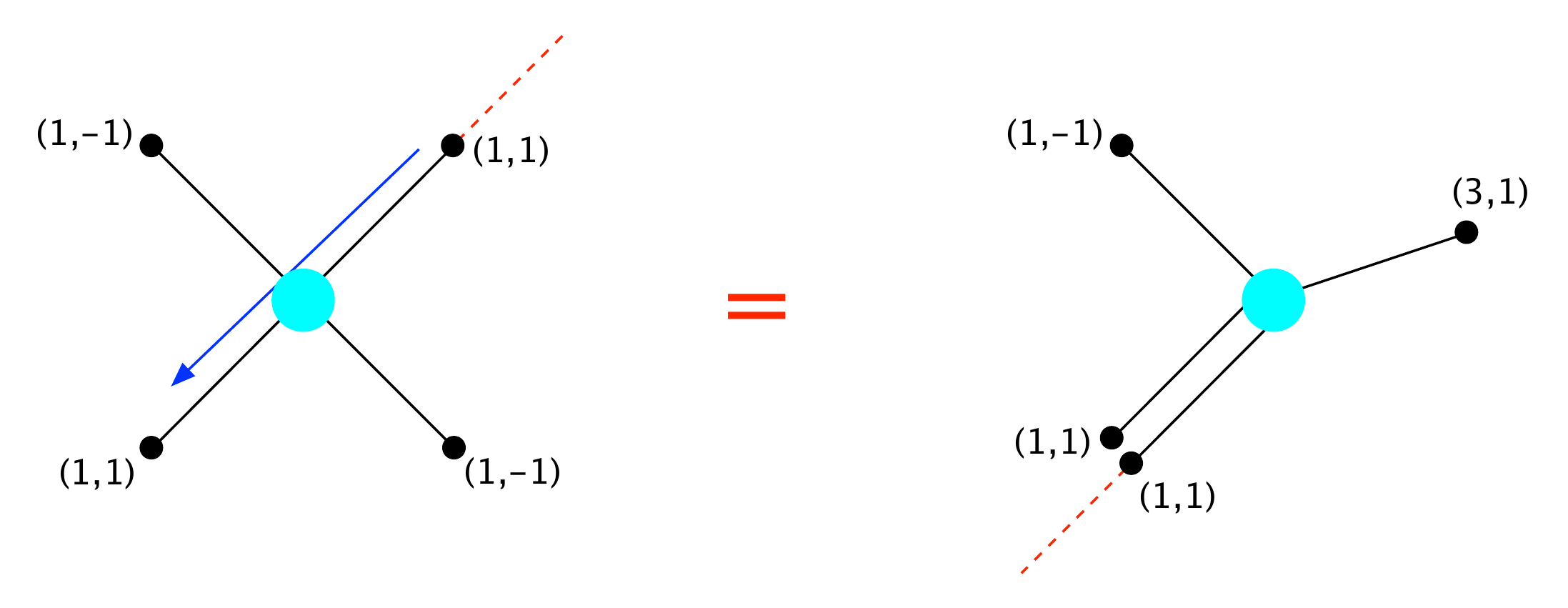} 
\caption{Equivalent 5-brane junctions for the rank 1 $E_1$ theory.}
\label{SimpleReduction3}
\end{figure}

The motion of a 7-brane along the 5-brane direction, on the other hand, has no effect on the field theory.
For example if we take the 7-brane to infinity along its 5-brane we just get the original infinite junction.
This continues to be true 
if we move the 7-brane towards the junction, shortening its 5-brane.
And it contiues to be true as the 7-brane crosses to the other side of the junction.
Three things happen in this case:
the 5-branes ending on the 7-brane reverse their orientation, 
additional 5-branes are created via the Hanany-Witten effect, some of which annihilate the 
orientation-reversed 5-branes, and other 7-branes and 5-branes
are transformed as we sweep the monodomy cut of the 7-brane to the other side.
The resulting 5-brane junction is different; it has a different set of $(p,q)$ charges.
But it describes the same 5d theory.
We refer to this as an {\em equivalent web}.
A simple example of this is shown in Fig.~\ref{SimpleReduction3}, which shows two 
equivalent 5-brane junctions describing the rank 1 $E_1$ theory.
Equivalences of this type are very useful in simplifying complicated looking 5-brane webs,
and in identifying 5d gauge theory phases when they exist.

\subsection{Generalized s-rule}

Importantly, the introduction of 7-brane endpoints to a 5-brane junction
also leads to an additional condition for supersymmetry. Heuristically, this is understood 
using the equivalent web construction described above. If, when a 7-brane crosses a junction 
the number of 5-branes created is less than the original number of 5-branes, we will be
left with 5-branes having the ``wrong" orientation, namely anti-5-branes.
The resulting configuration is clearly non-supersymmetric, and therefore so is the original one.
A special case of this condition is the {\em s-rule}:
\begin{description}
\item{\bf s-rule:} Given a $(p,q)$ 7-brane and an $(r,s)$ 5-brane, with both $(p,q)$ and $(r,s)$
being a pair of co-prime integers, the number of $(p,q)$ 5-branes that can be suspended 
supersymmetrically between them is at most $|ps - qr|$ 
\cite{Benini:2009gi}.\footnote{The {\em s-rule} was originally formulated for D3-branes
between a linked NS5-D5 pair in \cite{Hanany:1996ie}, and later derived for fundamental
strings between linked D-branes in \cite{Bachas:1997sc}. 
There were subsequent generalizations to D3-branes between $(p,q)$ 5-branes and $(r,s)$ 5-branes 
\cite{Kitao:1998mf}, and to $(p,q)$ strings 
between $(p,q)$ 7-branes and $(r,s)$ strings in \cite{Mikhailov:1998bx,Bergman:1998ej}.
The s-rule for $(p,q)$ 5-branes is identical to the one for $(p,q)$ strings.}
\end{description}
This follows immediately from a single Hanany-Witten move, in which $|ps-qr|$ 5-branes are created.

The s-rule provides a bound for a special class of triple-5-brane junctions.
One would like to find the generalization of this bound that applies to any multi-5-brane junction.
In other words, given an $n$-junction with charges $\{(p_i,q_i)\}$, that are not necessarily co-prime,
what is the condition on $\{(p_i,q_i)\}$ for it to be supersymmetric?

While the general condition for supersymmetry of 5-brane junctions has not yet been clearly formulated,
the analogous condition for the related system of string junctions has been known for a while
\cite{DeWolfe:1998bi,Iqbal:1998xb}.
Those studies were aimed at finding the spectrum of BPS states of four-dimensional ${\cal N}=2$ 
supersymmetric gauge theories living on 3-brane probes in 7-brane backgrounds.
The most general BPS state is described by a supersymmetric string web with some prongs ending on 3-branes,
and some ending on 7-branes.
The condition given in \cite{DeWolfe:1998bi,Iqbal:1998xb} is expressed in terms of a specific
$SL(2,\mathbb{Z})$ invariant quantity \footnote{We assume here that the charges $(p_i,q_i)$ are either 
all incoming or all outgoing, and that they are ordered either clockwise or counterclockwise 
around the junction.}
\be
\label{Intersection}
{\cal I} = \left|\sum_{1\leq i < j \leq n_7+n_3}
\mbox{det} \left(
\begin{array}{cc}
p_i & q_i \\
p_j & q_j
\end{array}
\right)\right|
- \sum_{i=1}^{n_7} \left(\mbox{gcd}(p_i,q_i)\right)^2 \,,
\ee
and is given by the inequality
\be
\label{IntersectionInequality}
{\cal I} \geq \mbox{} - 2 + \sum_{j=n_7+1}^{n_7+n_3} \mbox{gcd}(p_j,q_j) \,.
\ee
This condition on the multi-string-junction originates from its lift to M-theory,
where it corresponds to an M2-brane wrapping a holomorphic curve in an
elliptically fibered K3 surface.
The quantity ${\cal I}$ is the self-intersection number of the curve, and the inequality in 
(\ref{IntersectionInequality}) follows from the equality ${\cal I} = 2g -2 + b$,
the identification of the number of boundaries $b$ with the number of co-prime $(p,q)$-strings 
ending on 3-branes, and the non-negativity of the genus $g$.
The genus itself then corresponds to the number of possible faces in the web,
which for the 5-brane junction gives the dimension of the Coulomb branch of the 5d SCFT,
\be
d_C = \frac{{\cal I} + 2}{2}\,.
\ee

\begin{figure}[h!]
\center
\includegraphics[width=0.5\textwidth]{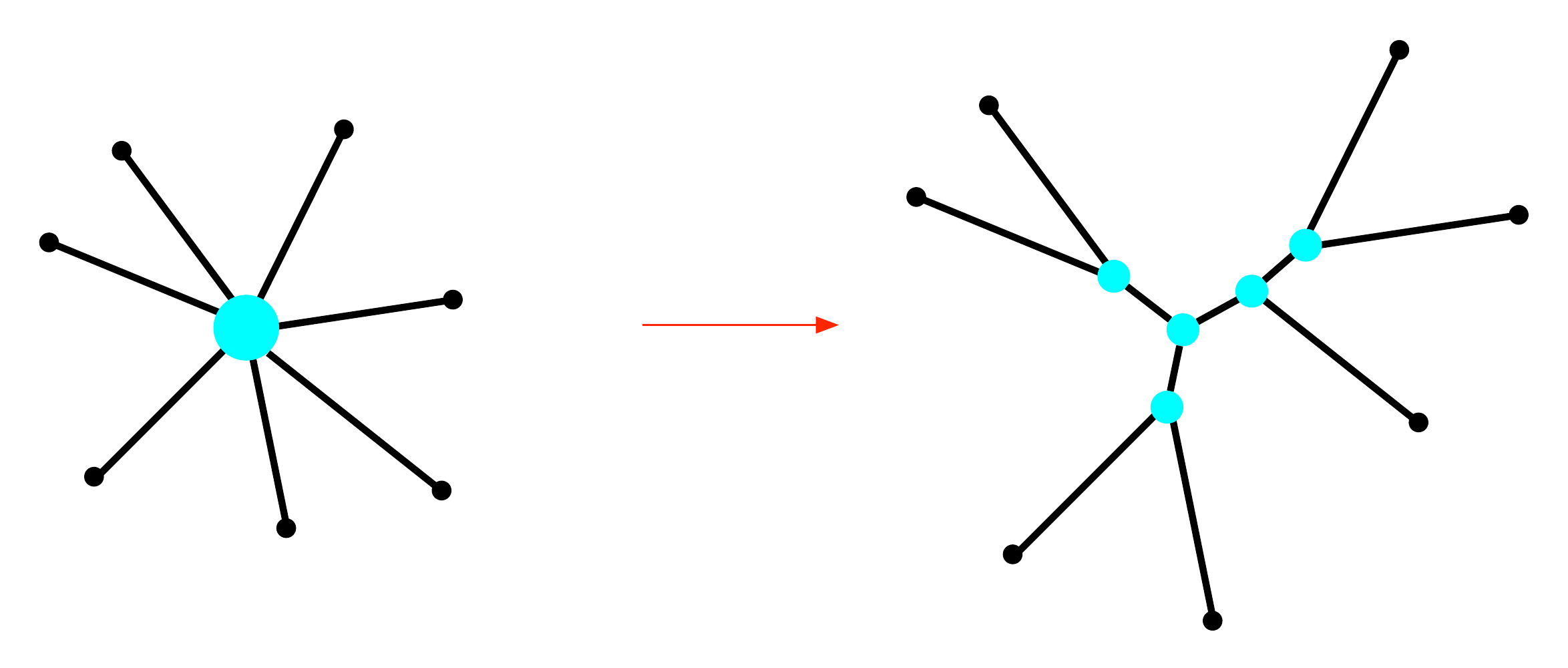} 
\caption{A generic mass deformation of an $n$-junction}
\label{GenericDeformation}
\end{figure}

Upon closer examination however, it appears that the above criterion is only applicable for
{\em irreducible} junctions, namely for junctions that cannot be separated into sub-junctions.
This issue should be investigated further.
For our purpose, however, it is sufficient to concentrate on the condition for 3-junctions.
If an $n$-junction is supersymmetric it should have an $(n-3)$-dimensional space of supersymmetric mass deformations.
At a generic point in this space the $n$-junction is deformed into a web consisting of 3-junctions 
and segments, Fig.~\ref{GenericDeformation}. 
Therefore a sufficient condition for the $n$-junction to be supersymmetric, is that it can be deformed
into a web consisting of 3-junctions and segments, where all the 3-junctions are supersymmetric.
For a 3-junction with charges $\{(p_1,q_1),(p_2,q_2),(-p_1-p_2,-q_1-q_2)\}$ 
the bound (\ref{IntersectionInequality}) becomes
\be
\label{Generalizedsrule}
{\cal I} = |p_1q_2 - p_2q_1| 
- \sum_{i=1}^{n_7} \left(\mbox{gcd}(p_i,q_i)\right)^2 \geq \mbox{} - 2 + \sum_{i=n_7 +1}^3 \mbox{gcd}(p_i,q_i) \,.
\ee
If the 3-junction is reducible, namely if it consists of multiple copies of the same basic 3-junction,
this bound applies to the basic 3-junction.
For $n_7=1$ and $(p_2,q_2)$ relatively prime this reduces to the s-rule.

\subsection{Trivial junctions}

\begin{figure}[h!]
\center
\includegraphics[width=0.7\textwidth]{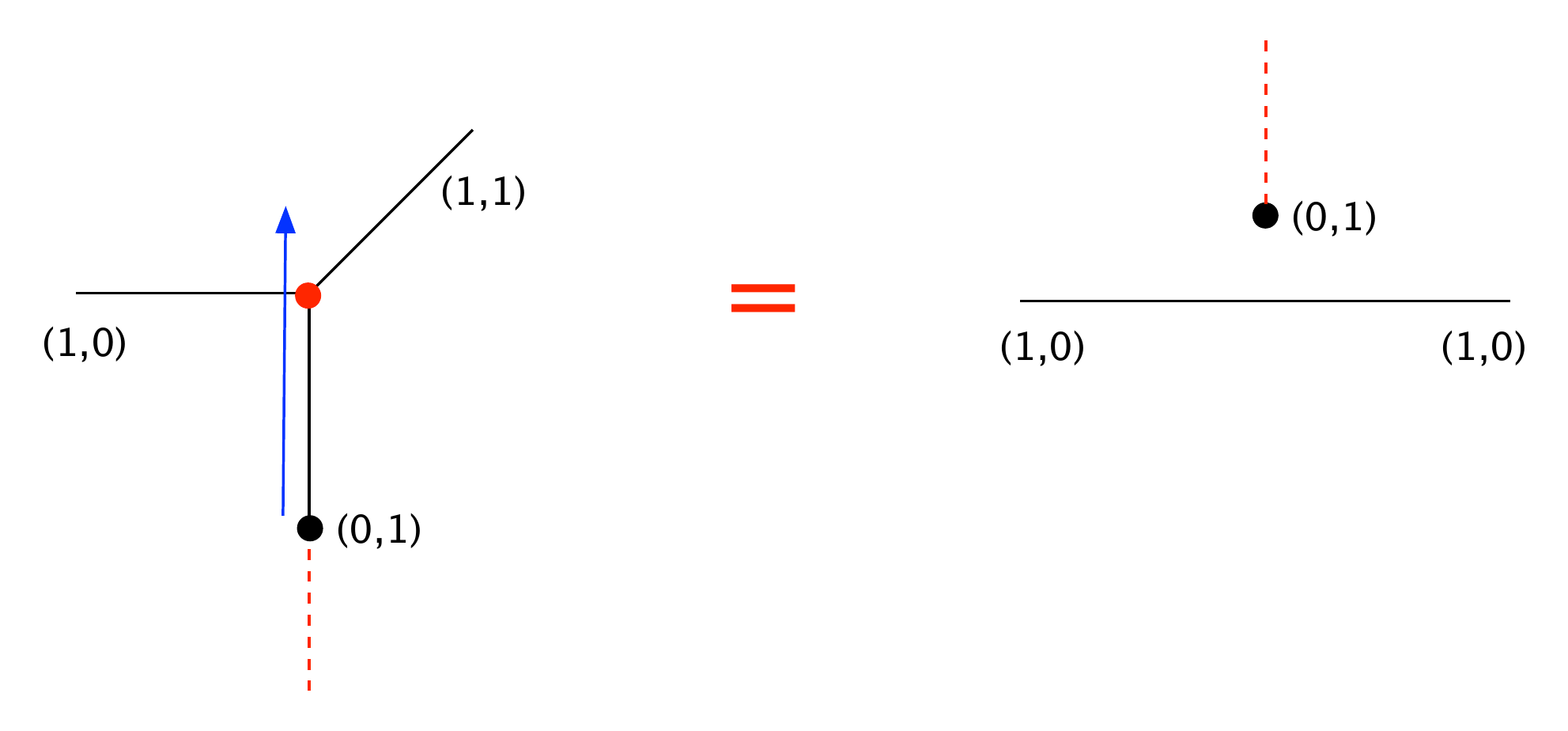} 
\caption{A simple trivial junction}
\label{SimpleReduction}
\end{figure}

\begin{figure}[h!]
\center
\includegraphics[width=0.8\textwidth]{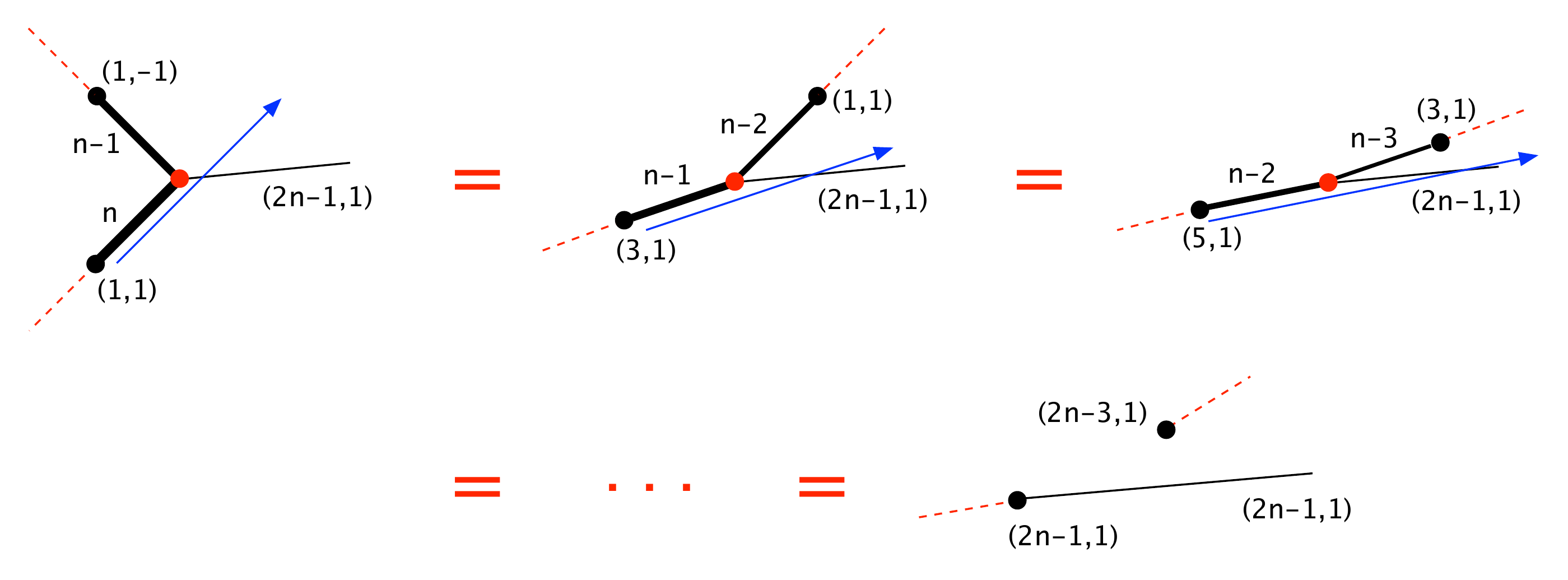} 
\caption{A compound trivial junction}
\label{SimpleReduction2}
\end{figure}

Hanany-Witten moves can also lead to a simpler configuration with a detached 7-brane.
This happens when the number of created 5-branes is equal to the original number of 
5-branes ending on the 7-brane.
For a 3-junction this eliminates the junction completely, leaving just a 5-brane and a detached 7-brane. We refer to this as a {\em trivial} junction, and mark it with a red circle.
A trivial 3-junction corresponds to a free hypermultiplet.
A simple example of a trivial junction is shown in Fig.~\ref{SimpleReduction}.
When the (0,1) 7-brane crosses the (1,0) 5-brane it loses its (0,1) 5-brane connection,
and we end up with a detached (0,1) 7-brane and a (1,0) 5-brane.
A trivial junction may require several Hanany-Witten moves to detach a 7-brane, as shown in the example of 
Fig.~\ref{SimpleReduction2}.
Trivial junctions have no local deformations, {\em i.e.} $g=0$, and therefore saturate the inequality (\ref{IntersectionInequality}).
More specifically the self-intersection number of a trivial junction is ${\cal I} = 0,-1$ or $-2$,
depending on whether it ends on one, two, or three 7-branes, respectively.

\begin{figure}[h!]
\center
\includegraphics[width=0.7\textwidth]{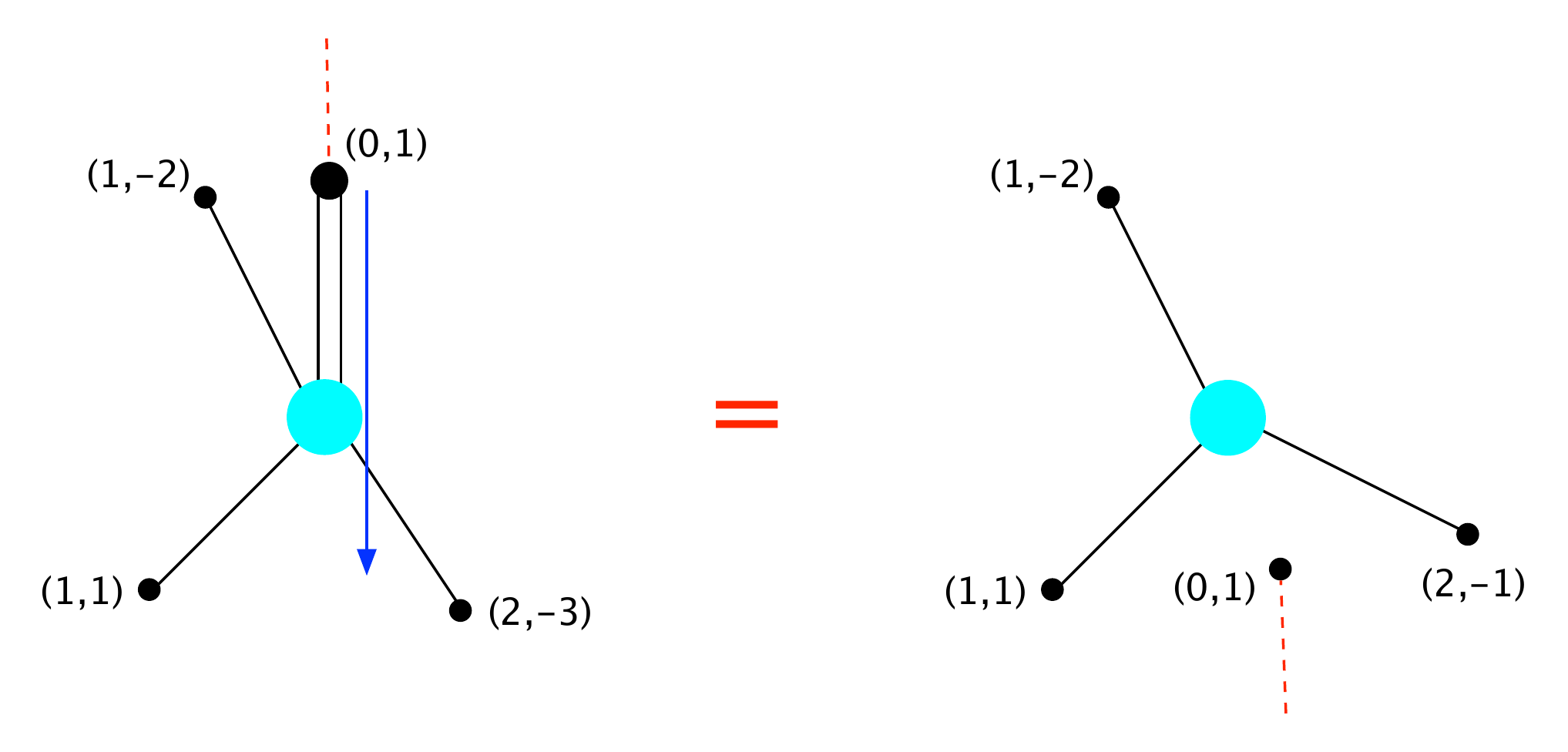} 
\caption{An example of a 1-trivial 4-junction describing the $E_0$ theory plus a free hypermultiplet.}
\label{OneTrivial4Junction}
\end{figure}

More generally, we define an $m$-trivial $n$-junction as an $n$-junction from which
$m$ 7-branes can be detached by Hanany-Witten moves.
Let us assume that the $n$-junction is irreducible, and that
all its legs end on 7-branes, namely
it corresponds to the full 5d SCFT.
Clearly $m \leq n-2$.
If $m=n-2$ the $n$-junction is equivalent to a 5-brane between two 7-branes and $n-2$
detached 7-branes. The detached 7-branes describe $n-2$ free hypermultiplets,
and the suspended 5-brane describes one more free hypermultiplet.
If $m<n-2$ the $n$-junction is equivalent to an $(n-m)$-junction plus $m$ detached 7-branes,
which describes a 5d SCFT given by the direct sum of the 5d SCFT described 
by the $(n-m)$-junction (which has a rank $n-m-3$ global symmetry) and $m$
free hypermultiplets.
An example of this with $n=4$ and $m=1$ is shown in Fig.~\ref{OneTrivial4Junction}.

\subsubsection{an alternative criterion for trivial junctions}

Above we identified trivial triple-junctions by their self-intersection number,
which is the minimal one preserving supersymmetry for the given number of 7-branes.
Here we will present an alternative, but equivalent, criterion.
The typical situation of interest involves a triple-junction with one internal leg
and two external legs ending on 7-branes. 
However the criterion is most easily formulated by imagining that the third leg also ends on a 7-brane.
If the charges of the first two 7-branes are $(p_1,q_1)$ and $(p_2,q_2)$,
then that of the third 7-brane is 
$(p_3,q_3) = (N_1p_1 + N_2p_2, N_1q_1,N_2q_2)$, where $N_1$ and $N_2$ are the multiplicities of the legs
ending on the first two 7-branes.
The junction is trivial if, by repeated moves of the first two 7-branes (the third 7-brane
is a spectator) one of them is detached.
This would leave a detached $(r,s)$ 7-brane and a $(p_3,q_3)$ 5-brane between two $(p_3,q_3)$ 7-branes: the other one of the two and the spectator 7-brane.
We assume here that $(p_3,q_3)$ are relatively prime.
The charge $(r,s)$ of the detached 7-brane depends on the original charges $(p_1,q_1)$, $(p_2,q_2)$,
and on the number of moves required for its detachment.
While the charges of the two active 7-branes change in this process, there are two 
useful invariant quantities.
The first is the total monodromy of the three 7-branes ordered counterclockwise, 
\be
M = M_{p_1,q_1} M_{p_2,q_2} M_{p_3,q_3}\,,
\ee
and the second is the Dirac pairing of the two active 7-branes,
\be
\Delta =
\mbox{det}\left(
\begin{array}{cc}
p_1 & q_2 \\
p_2 & q_2
\end{array}
\right) \,.
\ee
For a trivial junction these satisfy the relation
\begin{equation}
    {\rm Tr}\,M=2\,(1-\Delta^2)\,.
\end{equation} 
This can be seen by computing the monodromy and Dirac pairing in the detached configuration,
where
\be
M = M_{r,s} M_{p_3,q_3}^2 \,, \quad
\Delta =
\mbox{det}\left(
\begin{array}{cc}
r & s \\
p_3 & q_3
\end{array}
\right) \,.
\ee

As a consistency check, let us apply this criterion to the original junctions in 
Figs.~\ref{SimpleReduction}, \ref{SimpleReduction2}.
In the first example, adding a $(1,0)$ 7-brane and a spectator $(1,1)$ 7-brane, we have
$\mbox{Tr}(M) = \mbox{} -6$ and $\Delta = 2$, which satisfies the criterion.
In the second example, adding a spectator $(2n-1,1)$ 7-brane,
we also have $\mbox{Tr}(M) = \mbox{} -6$ and $\Delta = 2$, which again satisfies the criterion.

\end{document}